\documentclass[epj,nopacs]{svjour}
\usepackage{graphics}
\usepackage{graphics}
\usepackage{graphicx}
\usepackage{epsfig}
\usepackage{amssymb}
\usepackage{hyperref}
\usepackage{latexsym}
\usepackage{mathrsfs}
\usepackage{amsmath}
\usepackage{amscd}
\usepackage{color}
\usepackage{verbatim}
\usepackage[brazilian,english]{babel}
\begin{document}
\title{Numerical bifurcation analysis of two coupled FitzHugh-Nagumo oscillators}
\author{Anderson Hoff \and Juliana V. dos Santos \and Cesar
  Manchein\thanks{email: cesar.manchein@udesc.br} \and Holokx
  A. Albuquerque\thanks{email: holokx.albuquerque@udesc.br}}
\institute{Departamento de F\'\i sica, Universidade do Estado de Santa
  Catarina, 89219-710 Joinville, Brazil}
\date{Received: date / Revised version: date}
%
\abstract{The behavior of neurons can be modeled by the
  FitzHugh-Nagumo oscillator model, consisting of two nonlinear
  differential equations, which simulates the behavior of nerve
  impulse conduction through the neuronal membrane. In this work, we
  numerically study the dynamical behavior of two coupled
  FitzHugh-Nagumo oscillators. We consider unidirectional and
  bidirectional couplings, for which Lyapunov and isoperiodic diagrams
  were constructed calculating the Lyapunov exponents and the number
  of the local maxima of a variable in one period interval of the
  time-series, respectively. By numerical continuation method the
  bifurcation curves are also obtained for both couplings. The
  dynamics of the networks here investigated are presented in terms of
  the variation between the coupling strength of the oscillators and
  other parameters of the system. For the network of two oscillators
  unidirectionally coupled, the results show the existence of Arnold
  tongues, self-organized sequentially in a branch of a Stern-Brocot
  tree and by the bifurcation curves it became evident the connection
  between these Arnold tongues with other periodic structures in
  Lyapunov diagrams. That system also present multistability shown in
  the planes of the basin of attractions.
 }
\maketitle
%
\section{Introduction}
\label{introduction}
The FitzHugh-Nagumo (FHN) oscillator model \cite{Fitz,Nag} is a mathematical
description of the qualitative features of the nervous impulse
conduction in the neural cell, namely the membrane potential dynamics,
the resting bias of the excitable membrane and the existence of an
action potential threshold. The two-dimensional FHN oscillator is
described by the variables $u$ and $w$, that represent the potential
of the cell membrane performing the excitability of the system,
and the recovery variable after an action potential, respectively.
Following Campbell and Waite \cite{camp}, such model is represented
by the following differential equations
\begin{equation}\label{eqFHN1}
\begin{split}
\frac{du}{dt} &= c(w+u-\frac{1}{3}u^3+J(t)), \\
\frac{dw}{dt} &= -\frac{1}{c}(u-a+bw),
\end{split}
\end{equation}
where $J(t)$ is a time-dependent function that represents the external stimulus,
and $a$, $b$, and $c$ are parameters \cite{Fitz,Nag}.
Such system can produce chaotic motion only if an
external stimulus $J(t)$ is applied \cite{pank,zamb}. Oscillatory
chaotic behaviors can also be induced considering the coupling of
FHN oscillators in a network \cite{camp,yana,marze}.

Here, we consider two coupled FHN oscillators with two
types of coupling, namely unidirectional and bidirectional. Our main
interest is to obtain the bifurcation curves and consequently in
describing the dynamics of the variables regarding the
coupling strength and other system parameters in the parameter-planes.
The parameter-planes are diagrams
when two parameters of the system are varied and the other are kept
constant. The colors codify some quantity that can be computed on the
model. Usually, this quantity is the Lyapunov exponent
\cite{alb,alan,denis,med}, periods \cite{alan,gal1,gal2}, or
other invariant quantities \cite{alan,zou}. That procedure
allows us to identify regions of periodic and chaotic behavior, and
recently it is applied in several models \cite{alan,manc,gal4,stoop}.
A general feature is observed in these parameter-planes,
the existence of {\it shrimp}-shaped periodic structures embedded in
chaotic domains \cite{gal4,stoop}.

In this paper we investigate the dynamics of
two coupled FHN oscillators as a function of the type of coupling,
without external stimulus. The dynamics is investigated in the
parameter-planes, namely via Lyapunov and isoperiodic diagrams
\cite{alan,gal1,gal2,cris,manc1},
and we compare their structural changes as the type of coupling.
It is shown the limitation of the Lyapunov diagrams to distinguish the quasi-periodic
from periodic oscillations. For two FHN oscillators unidirectionally coupled we report periodicity
domains similar to the Arnold tongues. The role of the coupling in FHN-networks was studied
in recent papers \cite{lia,chiang,aqil}, but with different
approaches of those studied here. In those papers, the interesting is in the
synchronization phenomena, and in our work we carry out a
systematic dynamical study of the two FHN oscillators with respect to the coupling type.

This paper is organized as follows. In Section \ref{fit} the network
models are introduced, and in the Section \ref{num} we present the
numerical results concerning with each coupling type. In Section \ref{conc}
we present the conclusions of our work.

\section{FitzHugh-Nagumo network models}
\label{fit}

In this section we present the FHN-network models for two types of coupling
between two oscillators, namely unidirectional and bidirectional coupling.

Following Campbell and Waite \cite{camp}, for the unidirectional coupling
case between two oscillators, the FHN-network model is given by
\begin{equation}\label{eqFHN01}
\begin{split}
\frac{dx_1}{dt} &= c(y_1+x_1-\frac{1}{3}x_1^3)+\gamma(x_1-x_2), \\
\frac{dy_1}{dt} &= -\frac{1}{c}(x_1-a+by_1), \\
\frac{dx_2}{dt} &= c(y_2+x_2-\frac{1}{3}x_2^3), \\
\frac{dy_2}{dt} &= -\frac{1}{c}(x_2-a+by_2),
\end{split}
\end{equation}
where $x_i$ and $y_i$, $i = 1,2$, represent the voltage across the cell membrane,
and the recovery state of the resting membrane of a neuron,
respectively. On the other hand $a$, $b$, and $c$ are parameters,
and $\gamma$ is the coupling strength between the network elements.

For bidirectional coupling case between two oscillators, the FHN-network model can be written as
\begin{equation}\label{eqFHN3}
\begin{split}
\frac{dx_1}{dt} &= c(y_1+x_1-\frac{1}{3}x_1^3)+\gamma(x_1-x_2), \\
\frac{dy_1}{dt} &= -\frac{1}{c}(x_1-a+by_1), \\
\frac{dx_2}{dt} &= c(y_2+x_2-\frac{1}{3}x_2^3)+\gamma(x_2-x_1), \\
\frac{dy_2}{dt} &= -\frac{1}{c}(x_2-a+by_2).
\end{split}
\end{equation}
Variables and parameters have the same meaning as in system~(\ref{eqFHN01}).

In the next section we present and discuss the results obtained from the numerical
solutions of those equations whose behaviors depend on the four above-mentioned
parameters. The results are essentially presented in two-dimensional diagrams, using the
Lyapunov exponents and the periods as the invariant quantities, here namely
Lyapunov and isoperiodic diagrams, respectively, as reported previously \cite{gal1,gal2}.
Bifurcation curves obtained by numerical continuation method \cite{kuz} are also presented
for both network models.

The Lyapunov diagram plots are obtained for the following parameter
combinations: $\gamma \times a$ with $b$ and $c$ fixed, $\gamma \times b$
with $a$ and $c$ fixed, and $\gamma \times c$ with $a$ and $b$ fixed,
for the systems~(\ref{eqFHN01}) and~(\ref{eqFHN3}). They
were constructed using the Lyapunov exponents numerically
calculated for these systems. To evaluate the Lyapunov spectra,
the Eqs.~(\ref{eqFHN01}) and~(\ref{eqFHN3}) are
numerically solved by the Runge-Kutta method with fixed time step in $1.0
\times 10^{-1}$ and $5.0 \times 10^5$ integration steps, via the algorithm
proposed in Ref. \cite{wolf} for each parameter pair discretized in
a grid of $500 \times 500$ values. We
performed other tests using smaller time steps and the results are
practically unchanged. Therefore, we obtain $2.5 \times 10^5$ Lyapunov
spectra for each two-dimensional diagrams, where each spectrum has the number
of values of Lyapunov exponents equal to the number of dimensions of the network.

The isoperiodic diagram plots are also computed for the two coupled FHN oscillator models.
To evaluate the period for each time-series with a fixed set of parameters, we use the Runge-Kutta
method again, but now with variable time step and removed a transient time equal to
$5 \times 10^6$ and $1 \times 10^6$ integration steps to find the period
obtained by the maxima of the time-series with a precision
of $1 \times 10^{-3}$. The parameter pairs were
discretized in a grid of $10^3 \times 10^3$ values.

To obtain the bifurcation curves in the parameter-planes, we used
the numerical continuation method. For this purpose, the MATCONT
package \cite{kuz} was used, and Hopf, saddle-node, Neimark-Sacker, and
period-doubling curves were obtained unveiling the bifurcation structures of
the systems~(\ref{eqFHN01})~and~(\ref{eqFHN3}). The bifurcation curves are used
extensively to unveil the bifurcation structures of dynamical systems \cite{bar1,bar2,bar3,ray,gen,hoff}.
In one of the first papers on this matter \cite{bar3}, Barrio {\it et al.}
performed a detailed study of bifurcations
in the three-parametric three-dimensional R\"ossler model. Among other results,
the authors overlapped bifurcation curves, obtained by numerical continuation
method \cite{kuz}, on two-dimensional parame\-ter-spaces of the model, and some
periodic structures might be described by those curves, as also be done
recently in Ref. \cite{hoff}, for a four-dimensional Chua model.

\begin{figure*}[htb]
  \centering
  \includegraphics*[width=0.9\linewidth]{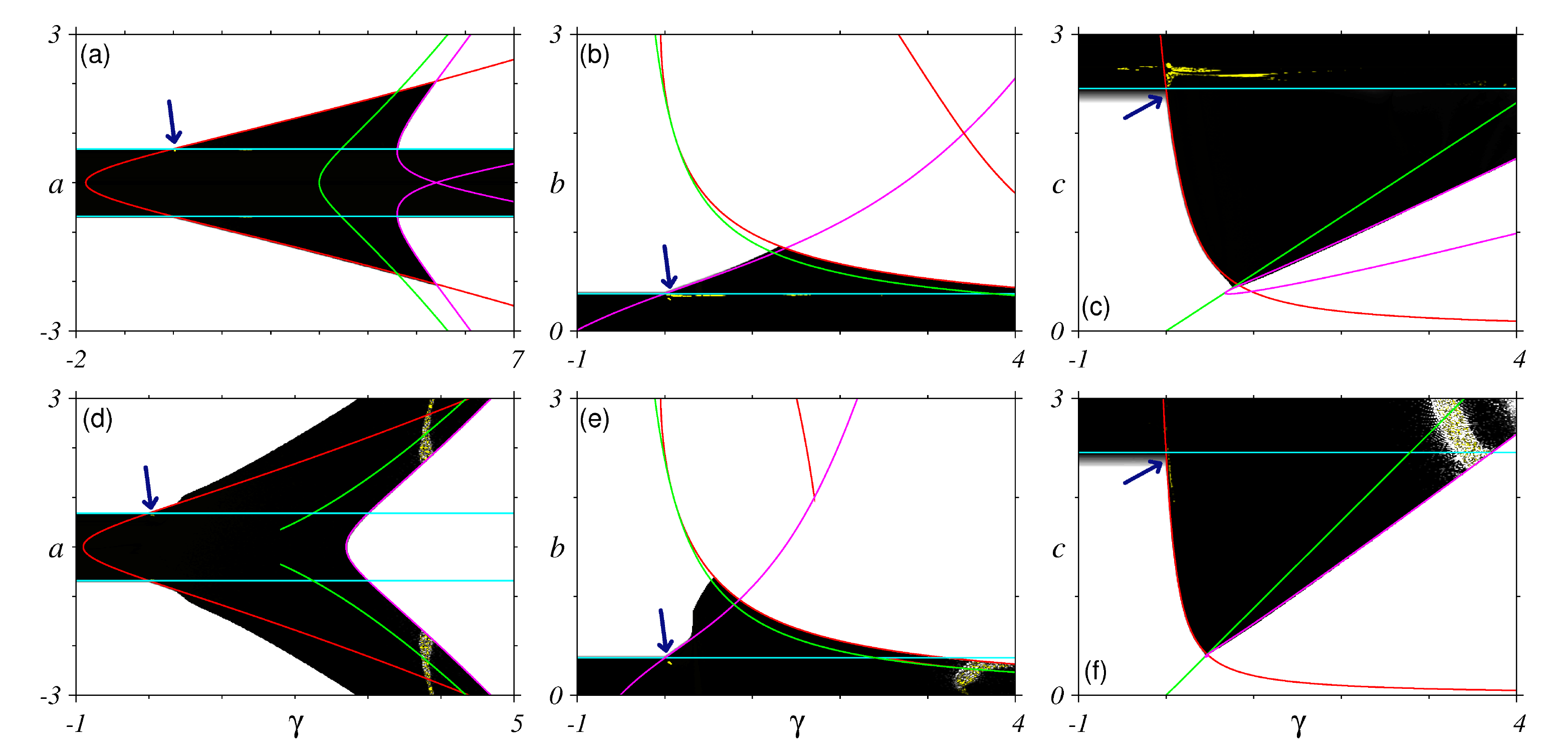}
  \caption{Bifurcation curves in the parameter-planes for two FHN oscillators
  with unidirectional coupling, system~(\ref{eqFHN01}) [top line (a)-(c)], and for
  two FHN oscillators with bidirectional coupling, system~(\ref{eqFHN3})
  [bottom line (d)-(f)]. (a) $\gamma \times a$ plane, with $b = 0.4$
  and $c = 2.0$. (b) $\gamma \times b$ plane with $a = 0.7$ and $c = 2.0$.
  (c) $\gamma \times c$ plane with $a = 0.7$ and $b = 0.4$.
  (d) $\gamma \times a$ plane, with $b = 0.4$ and $c = 2.0$.
  (e) $\gamma \times b$ plane with $a = 0.7$ and $c = 2.0$.
  (f) $\gamma \times c$ plane with $a = 0.7$ and $b = 0.4$.
  The colors white, black, and yellow represent the largest Lyapunov
  exponent and correspond to equilibrium points, periodic and chaotic
  motion, respectively. The arrows indicate small regions that present
  chaotic motion and are enlarged in Fig.~\ref{fig2}.}
  \label{fig1}
\end{figure*}

\section{Numerical results}
\label{num}

We begin our numerical investigation analyzing the para\-meter-planes of systems~(\ref{eqFHN01})
and~(\ref{eqFHN3}) in a large range of parameter values. The main goal
here is to know how is the dynamical behavior of the systems regarding
the bifurcation curves and Lyapunov exponents in those parameter values.
Based in the results presented in Ref.~\cite{camp}, we show in
Fig.~\ref{fig1} the parameter-planes of the systems~(\ref{eqFHN01}) ((a)-(c))
and~(\ref{eqFHN3}) ((d)-(f)) for the largest Lyapunov exponent with
bifurcation curves overlapped. The set of initial conditions used
is $(x_1, y_1, x_2, y_2) = (-0.1,0.1,0.5,-0.3)$, and the fixed parameters used
are $(a,b,c) = (0.7,0.4, 2.0)$, as the case. The cyan, red and magenta curves correspond
to Hopf bifurcations and the green curve corresponds to limit-point bifurcation.
The white, black and yellow regions correspond to equilibrium points, periodic
and chaotic motions, respectively, and codify the sign of the largest Lyapunov
exponent of the spectrum. For example, white corresponds to negative,
black to null and yellow to positive largest exponent. In Fig.~\ref{fig1}
the Hopf bifurcations occur when an equilibrium point
(white regions) lose its stability and a limit cycle (black regions) born.
This behavior is corroborated by the largest Lyapunov exponents, white
and black regions in the diagrams of Fig.~\ref{fig1}. As can also be
observed, the chaotic behaviors (yellow regions) are small regions in
the expanded diagrams and the arrows indicate where those regions are localized.

\begin{figure*}[htb]
  \centering
  \includegraphics*[width=0.9\linewidth]{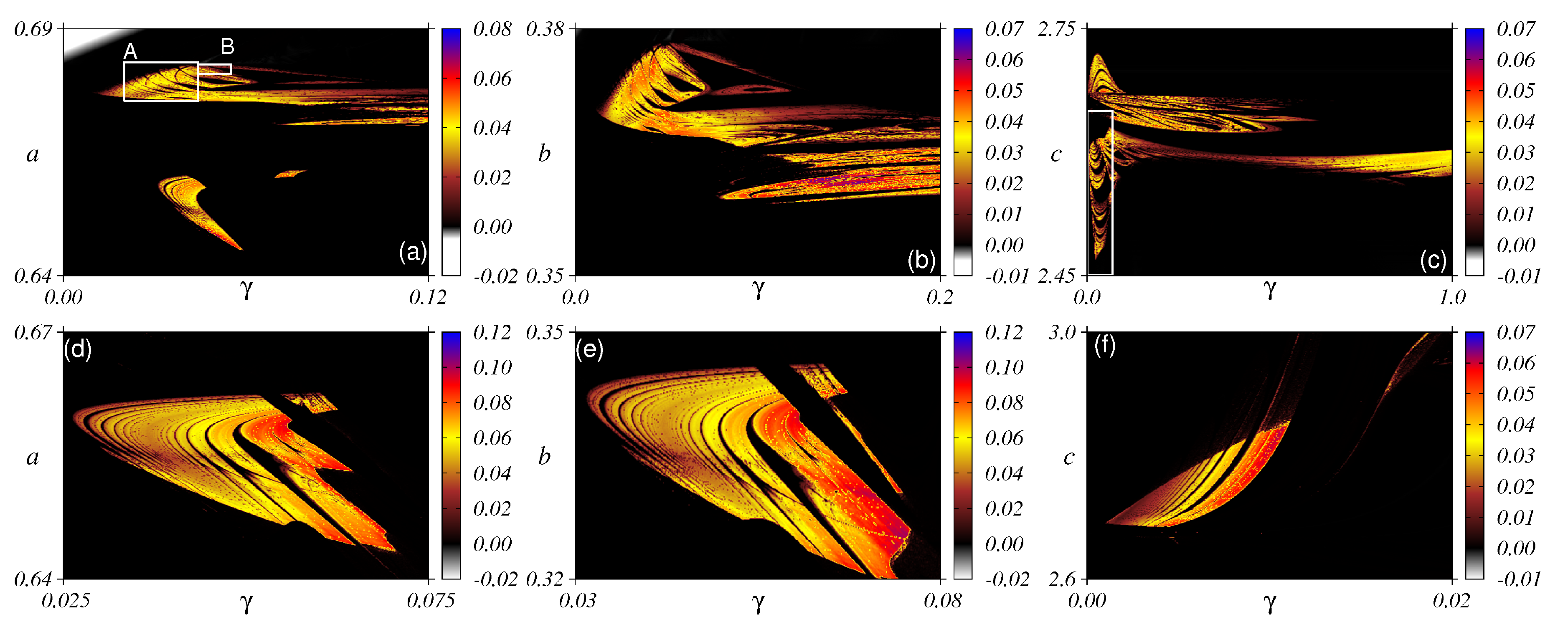}
  \caption{Lyapunov diagrams for the largest exponent of the spectra
  codified by colors, as the color bar at the right side, for two FHN
  oscillators with unidirectional coupling, system~(\ref{eqFHN01})
  [top line (a)-(c)], and for two FHN oscillators with bidirectional
  coupling, system~(\ref{eqFHN3}) [bottom line (d)-(f)]. The diagrams
  are amplifications of small regions located by the arrows in
  Fig.~\ref{fig1}. Periodic/quasi-periodic domains are represented by
  black colors, and chaotic ones are yellow-red-blue colors. The boxes
  in panels (a) and (c) delimit the regions magnified in
  Figs.~\ref{fig4} and~\ref{fig6}, respectively.}
  \label{fig2}
\end{figure*}

The equations used to describe the dynamical behavior of the
FHN-network with two unidirectional coupled oscillators are given by the
system~(\ref{eqFHN01}). For the same set of initial conditions used
in Fig.~\ref{fig1}, we show in Figs.~\ref{fig2}(a)-(c) the Lyapunov diagrams for
the following planes: (a) $\gamma \times a$ with $b$ and $c$ fixed,
(b) $\gamma \times b$ with $a$ and $c$ fixed, and (c) $\gamma \times c$ with $a$
and $b$ fixed. Such diagrams are amplifications of the small regions indicated by
the arrows in Figs.~\ref{fig1}(a)-(c), and are constructed with the largest
Lyapunov exponent of the spectrum codified in colors. For example, in
Fig.~\ref{fig2}, yellow-red-blue colors identify chaotic behavior, since
for points taken from these regions the largest exponent of the spectrum is
positive. In Figs.~\ref{fig2}(d)-(f) we present the Lyapunov diagrams for the largest
exponents of the spectra for two bidirectionally coupled models,
described by system~(\ref{eqFHN3}). Such diagrams are amplifications of the small
regions indicated by the arrows in Figs.~\ref{fig1}(d)-(f). As the
unidirectional case, the Lyapunov spectrum has four exponents, and
the diagrams shown in Figs.~\ref{fig2}(d)-(f) were constructed with the
largest one. The construction and interpretation of colors
follow the same of Figs.~\ref{fig2}(a)-(c).

All diagrams in Fig.~\ref{fig2}(a)-(f) show large periodic regions (regions in black)
and chaotic windows (yellow-red-blue regions) with periodic
structures (in black) embedded on them. As far as our knowledge, this feature is commonly
found in almost every nonlinear dissipative dynamical
systems~\cite{alb,alan,denis,med,gal1,gal2,zou,manc,gal4,stoop,cris,manc1,rec1}. Fig.~\ref{fig2}
show the existence of paths crossing the chaotic regions in which the
variables $x_i$, and $y_i$ have periodic behaviors. Moreover,
there exist regions in which these variables have chaotic
oscillations, depending of the parameter combinations. It is
important to observe that for the single FHN oscillator without external
forcing, namely a single neuron without external stimulus,
system~(\ref{eqFHN1}), the variables can not present chaotic motion.

\begin{figure*}[htb]
  \centering
  \includegraphics*[width=0.45\linewidth]{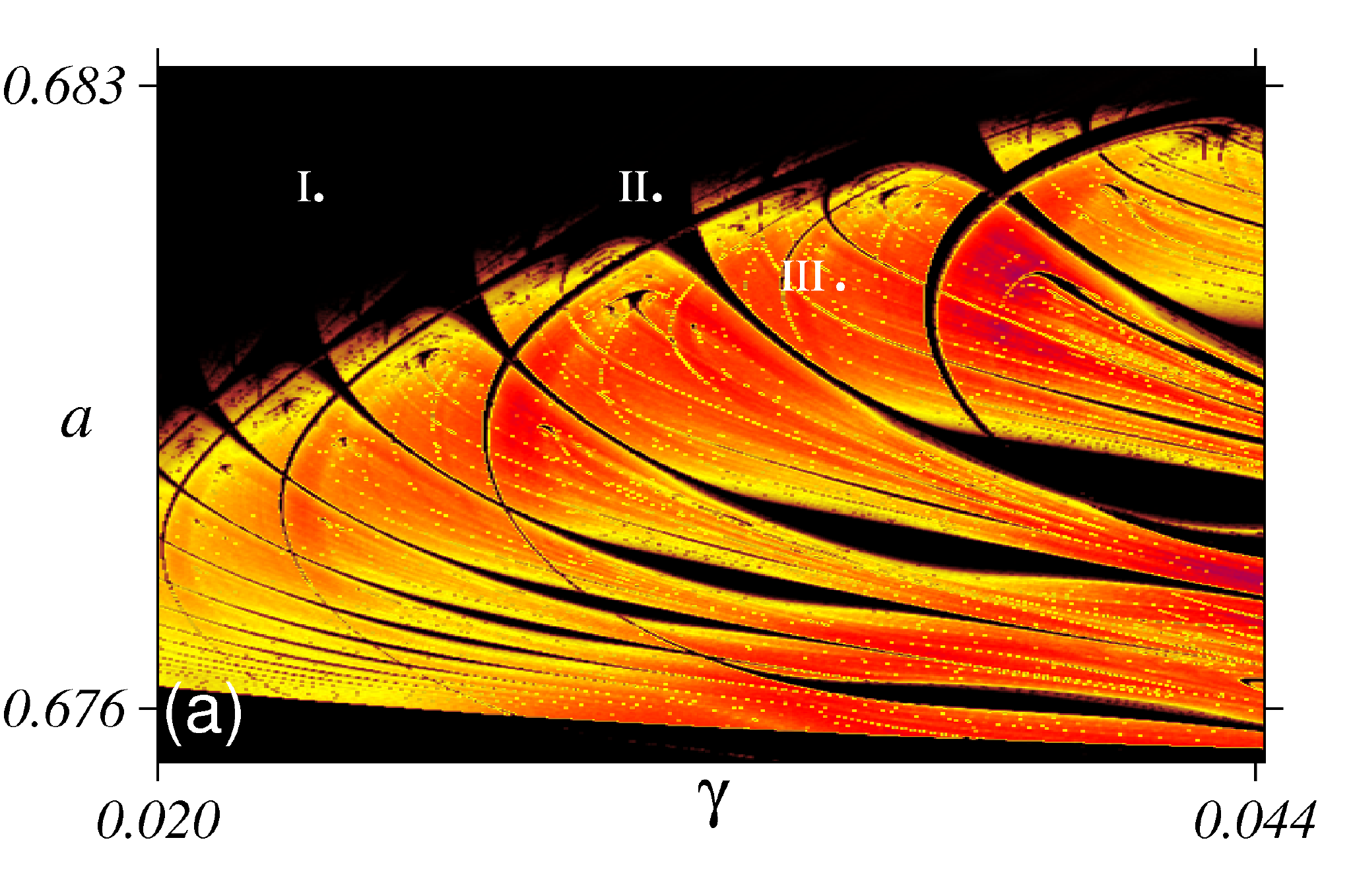}
  \includegraphics*[width=0.46\linewidth]{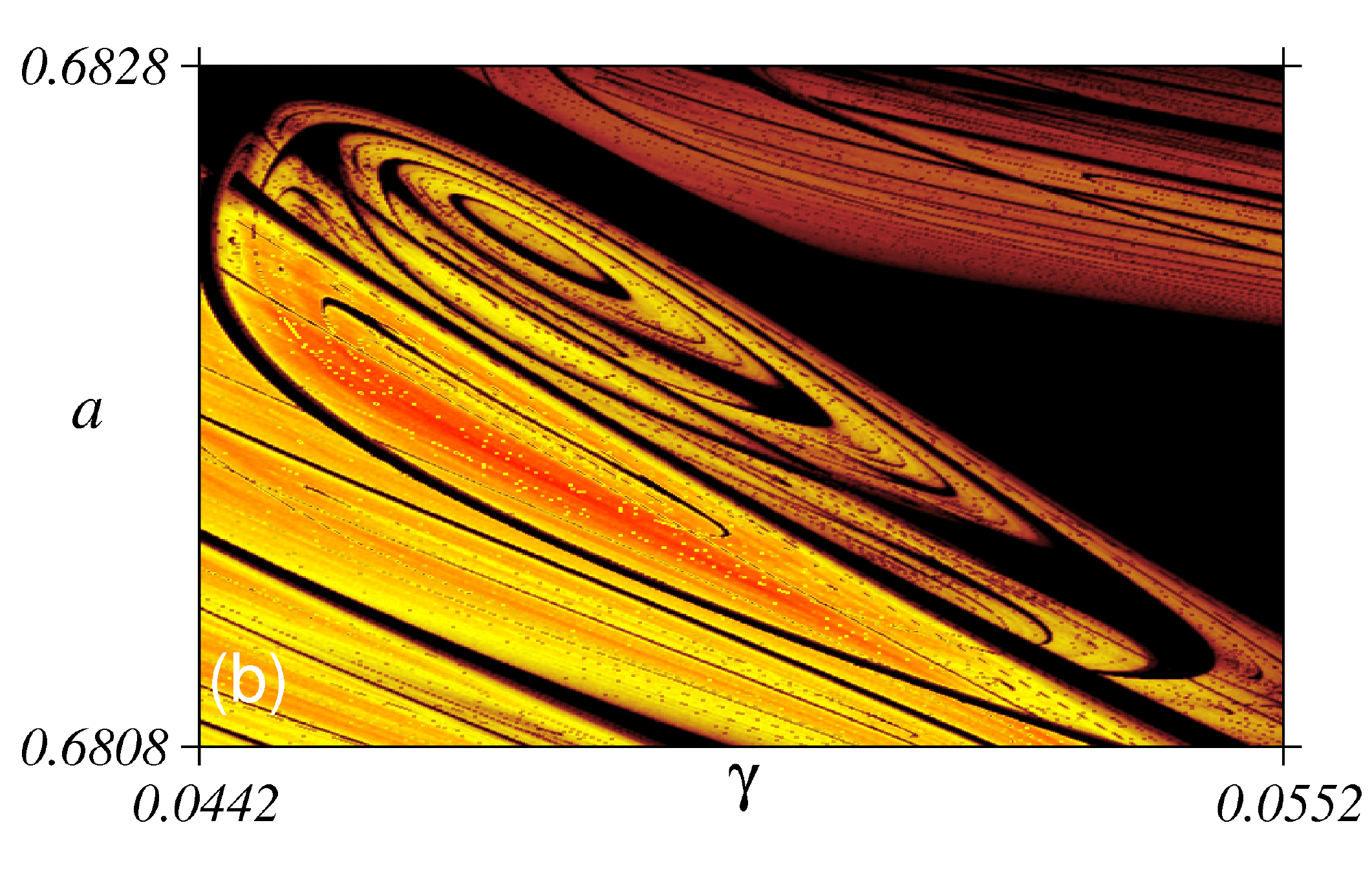}
  \includegraphics*[width=0.45\linewidth]{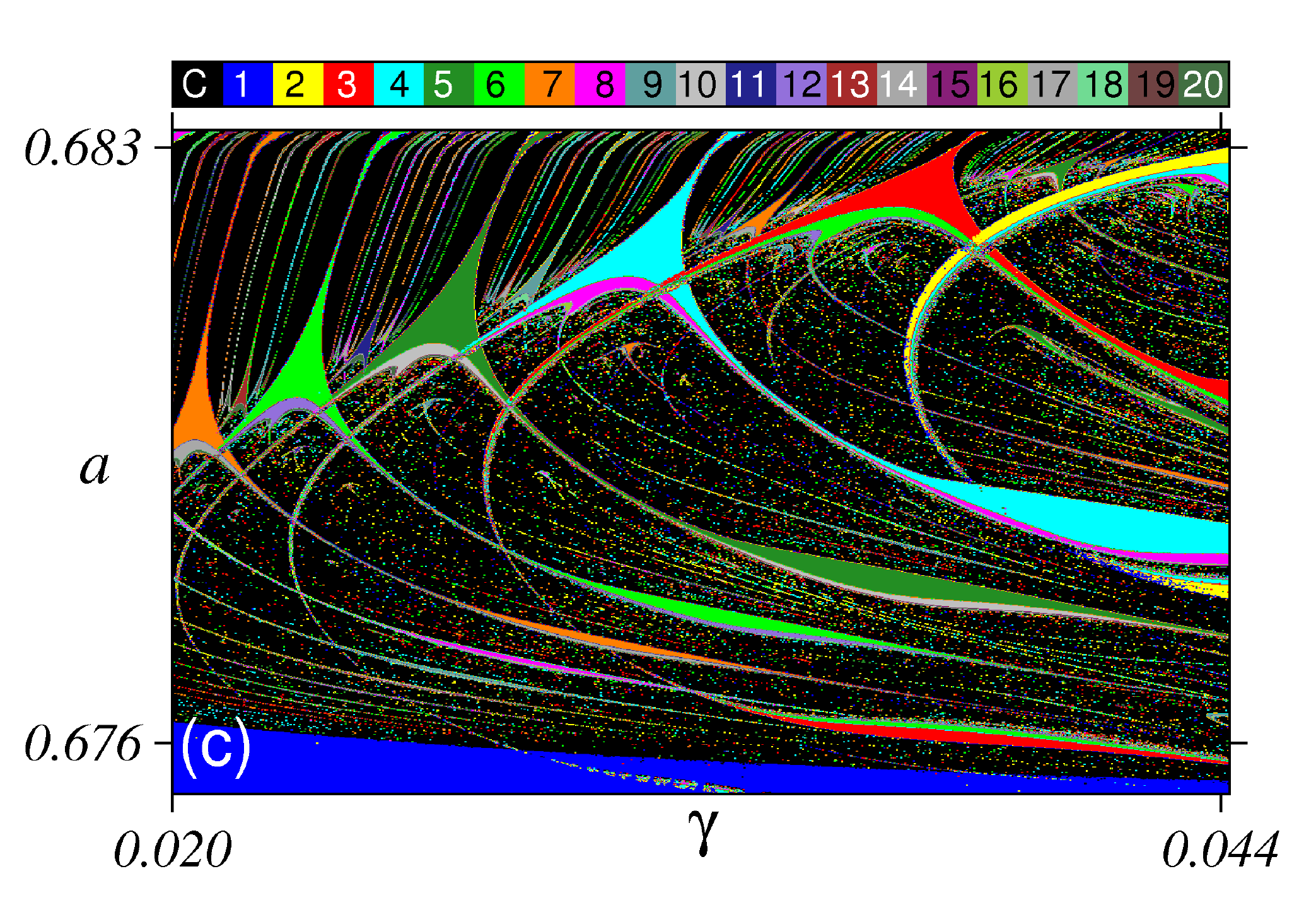}
  \includegraphics*[width=0.46\linewidth]{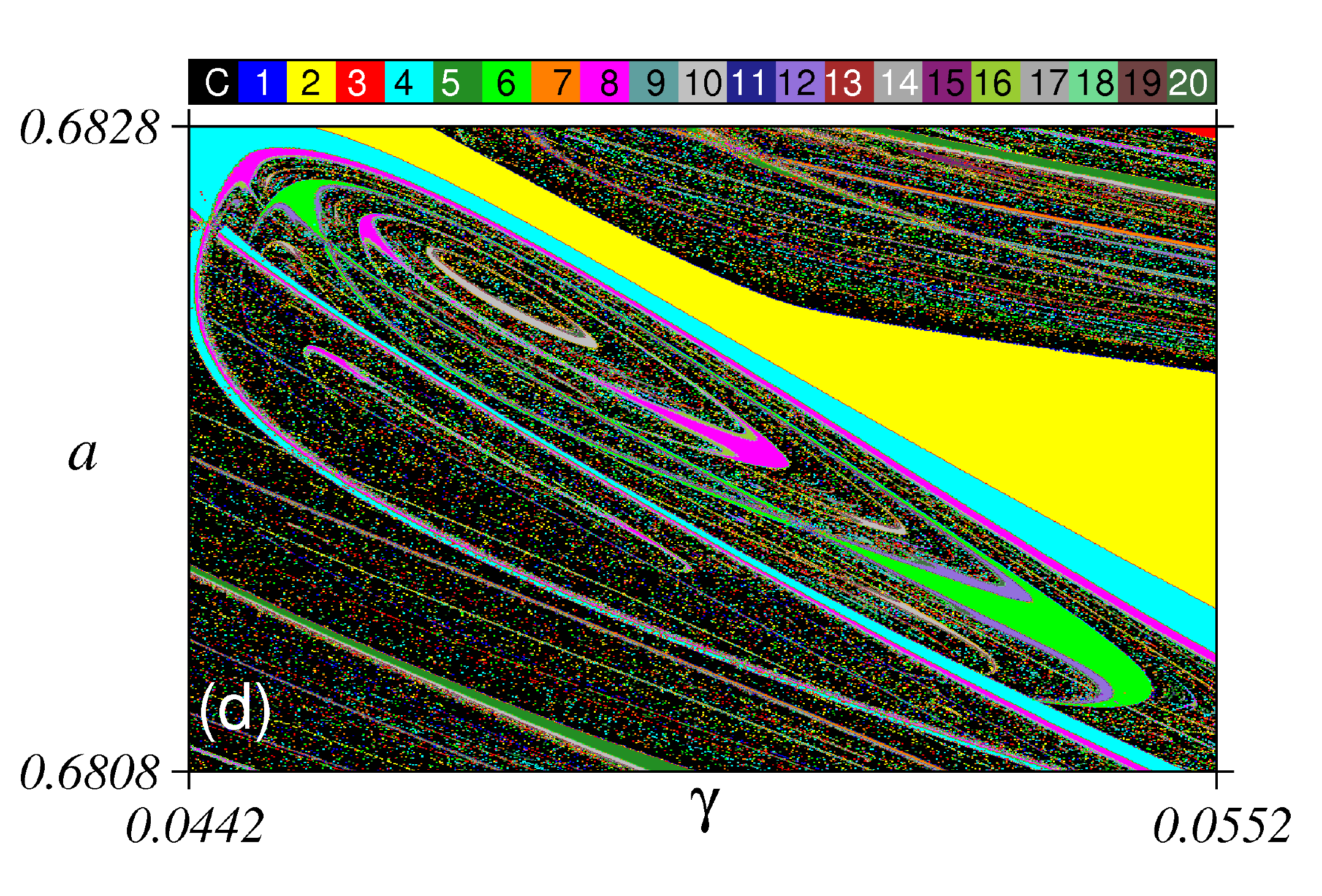}
  \caption{Unidirectional coupling case: (a) and (b) Lyapunov diagrams
  for the largest exponents, and
  (c) and (d) isoperiodic diagrams, both for the boxes A and B, respectively,
  in Fig.~\ref{fig2}(a). In (a) and (b) the black-yellow-red colors codify
  the largest Lyapunov exponent, and in (c) and (d) the top color bars codify
  the periods. Black corresponds to quasi-periodic/chaotic behaviors or periods
  greater than $20$.}
  \label{fig4}
\end{figure*}
\begin{figure*}[htb]
  \centering
  \includegraphics*[width=0.31\linewidth]{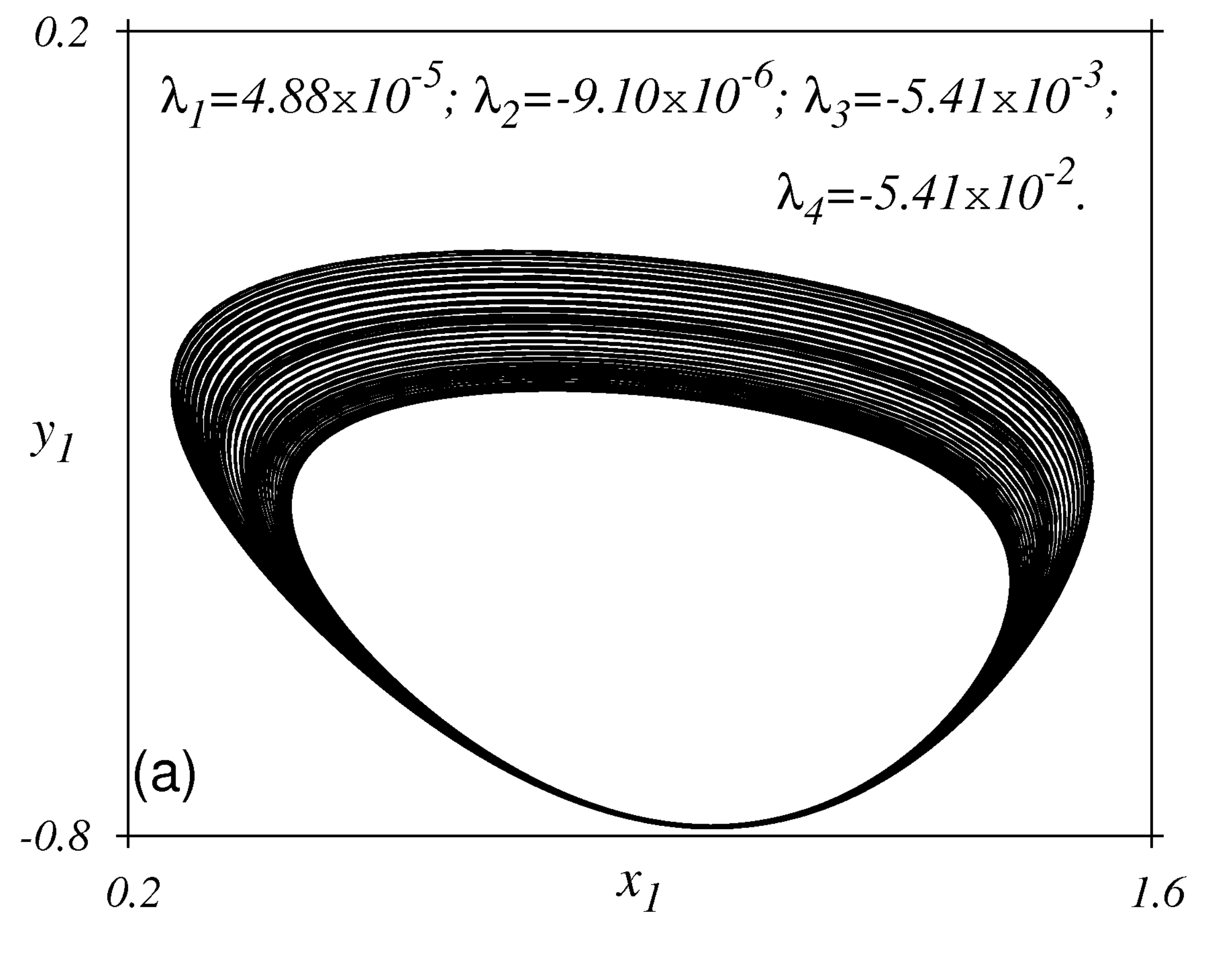}
  \includegraphics*[width=0.31\linewidth]{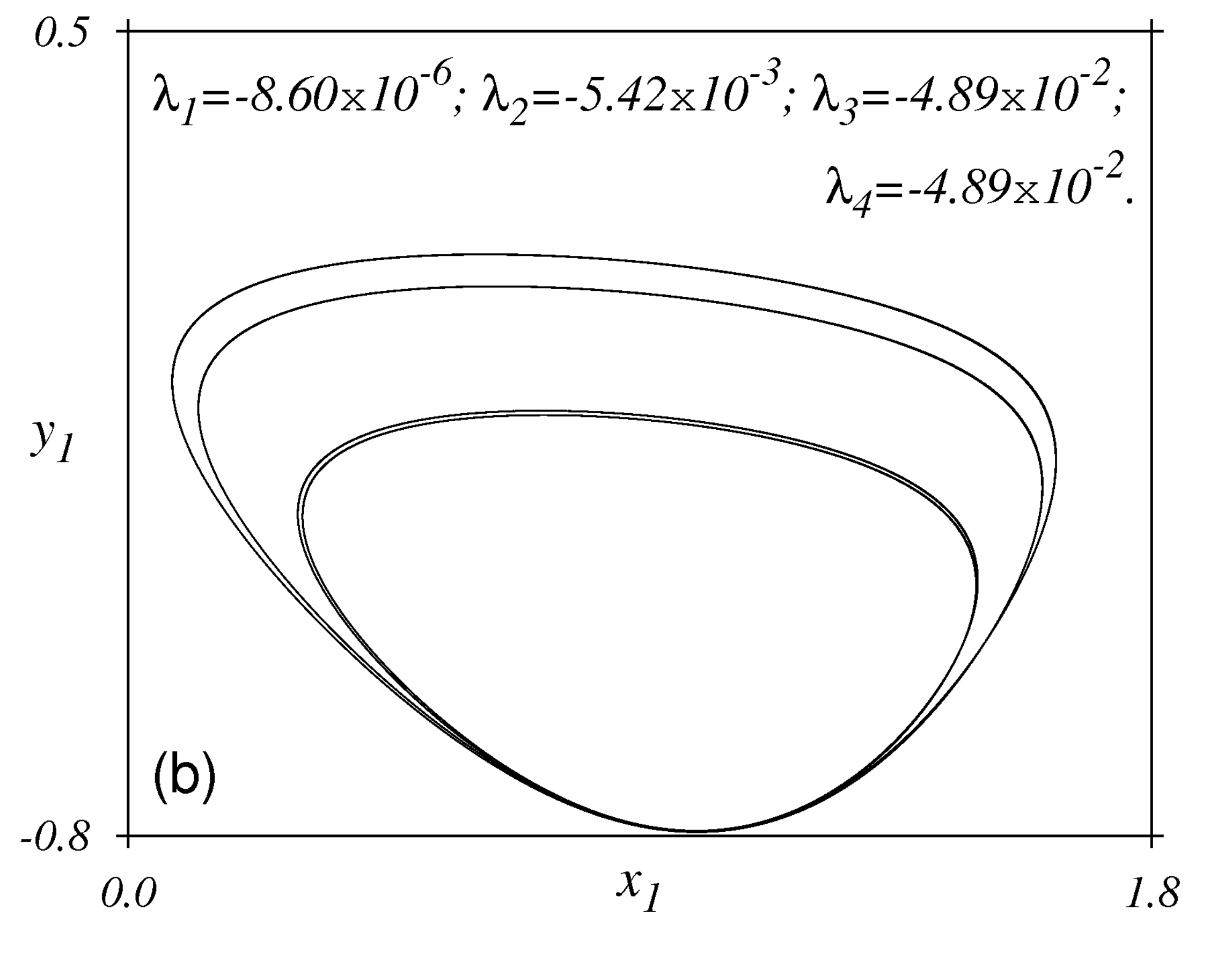}
  \includegraphics*[width=0.31\linewidth]{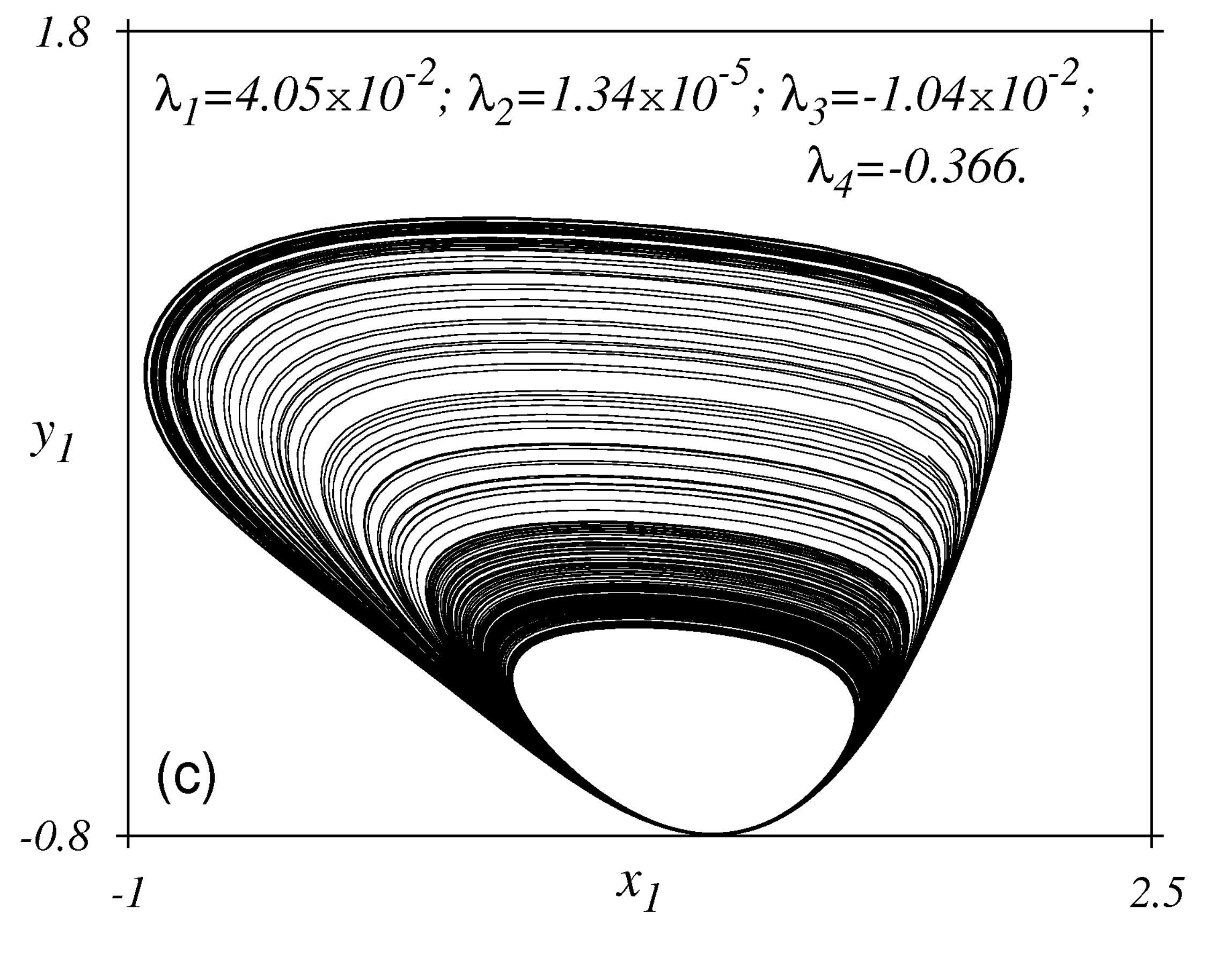}
  \caption{Attractors of three distinct points in Fig.~\ref{fig4}(a),
  showing (a) Torus-2 (point I), (b) periodic (point II), and
  (c) chaotic (point III) behaviors. The Lyapunov spectrum for each attractor
  is also shown.}
  \label{fig5}
\end{figure*}

The limitations of the Lyapunov diagram shown in Fig.~\ref{fig2} are
related to the distinction between periodic (in black) and chaotic
(yellow-red-blue) regions, and to observe the presence of periodic structures,
as cuspidal, non-cuspidal and shrimp-like structures~\cite{alan,manc1}.
In such diagrams is not possible to infer the periods of the structures
and their organization rules, as period-adding cascades~\cite{gal1,gal2,gal4},
and Stern-Brocot trees~\cite{gal1,gal2,gal4}.
Following recent works about isoperiodic diagrams~\cite{gal1,gal2,gal4},
that are parameter-planes using the numbers of spikes of the time series
(periods), we also constructed these diagrams for regions inside the boxes of
Figs.~\ref{fig2}(a) and (c).

Figure~\ref{fig4} shows the Lyapunov and the isoperiodic diagrams for the
boxes A and B in Fig.~\ref{fig2}(a).
In these diagrams it is clear the limitations of the Lyapunov diagram in both cases.
The isoperiodic diagram in Fig.~\ref{fig4}(c) unveils hidden regions and pieces of periodic
structures that are not perceived in Fig.~\ref{fig4}(a). For example, on the left top in Fig.~\ref{fig4}(a)
we observe a large black region with branches of periodic structures coming out
from it and extending over the chaotic region (yellow-red-blue region), however,
in Fig.~\ref{fig4}(c) we observe in this same region a combination of quasi-periodic behavior
(black color) with branches of periodic structures embedded on it. Moreover,
the organization pattern of these periodic structures delimits the quasi-periodic
region, upper black region in Fig.~\ref{fig4}(c), with the chaotic
region, lower black region in Fig.~\ref{fig4}(c). This organization
pattern is similar to the Arnold's tongues~\cite{schus,rech2}, and its
periodicity is similar to branches of Stern-Brocot tree observed in a
variety of continuous-time
systems~\cite{gal1,gal2,gal4,podh}. Basically, in Fig.~\ref{fig4}(c) the
organization rule of the periods can be summarized as follows. The primary
structures follow a period-adding cascade:
$7 \rightarrow 6 \rightarrow 5 \rightarrow 4 \rightarrow 3 \rightarrow 2$.
The sum of the periods of two consecutive primary structures is equal of the
period of secondary structures that are between these two primary structures.
For example: $7 + 6 \rightarrow 13, 6 + 5 \rightarrow 11, 5 + 4
\rightarrow 9, 4 + 3 \rightarrow 7, 3 + 2 \rightarrow 5$
(see Fig.~\ref{fig4}(c)). The sum of the period of a primary structure
with the consecutive secondary structure is equal of the period of
tertiary structure between these primary and secondary structures. For example:
$7 + 13 \rightarrow 20, 6 + 13 \rightarrow 19, 6 + 11 \rightarrow 17,
5 + 11 \rightarrow 16, 5 + 9 \rightarrow 14, 4 + 9 \rightarrow 13, 4 +
7 \rightarrow 11, 3 + 7 \rightarrow 10, 3 + 5 \rightarrow 8, 2 + 5
\rightarrow 7$.

\begin{figure*}[htb]
  \centering
  \includegraphics*[width=0.95\columnwidth]{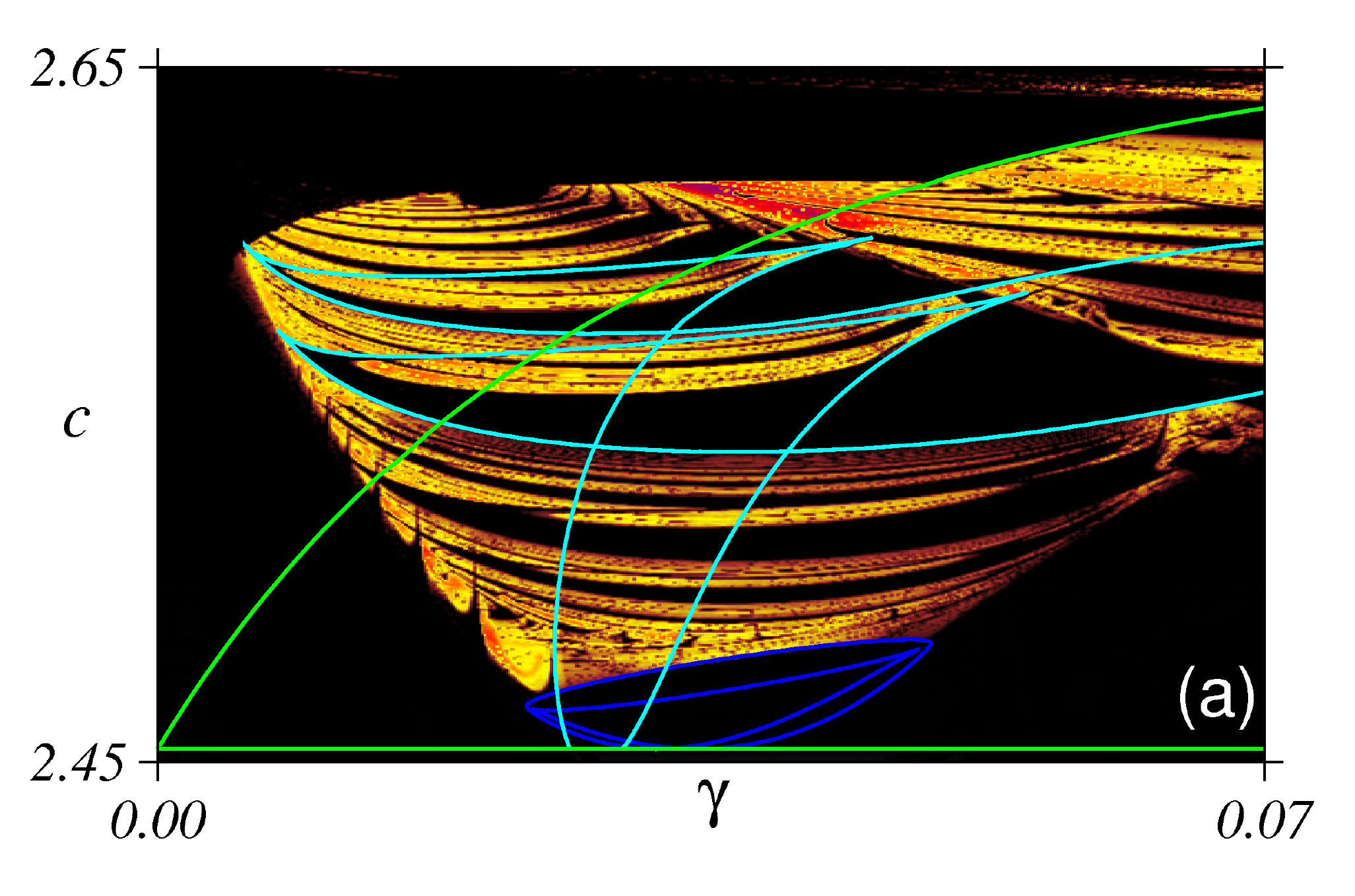}
  \includegraphics*[width=0.95\columnwidth]{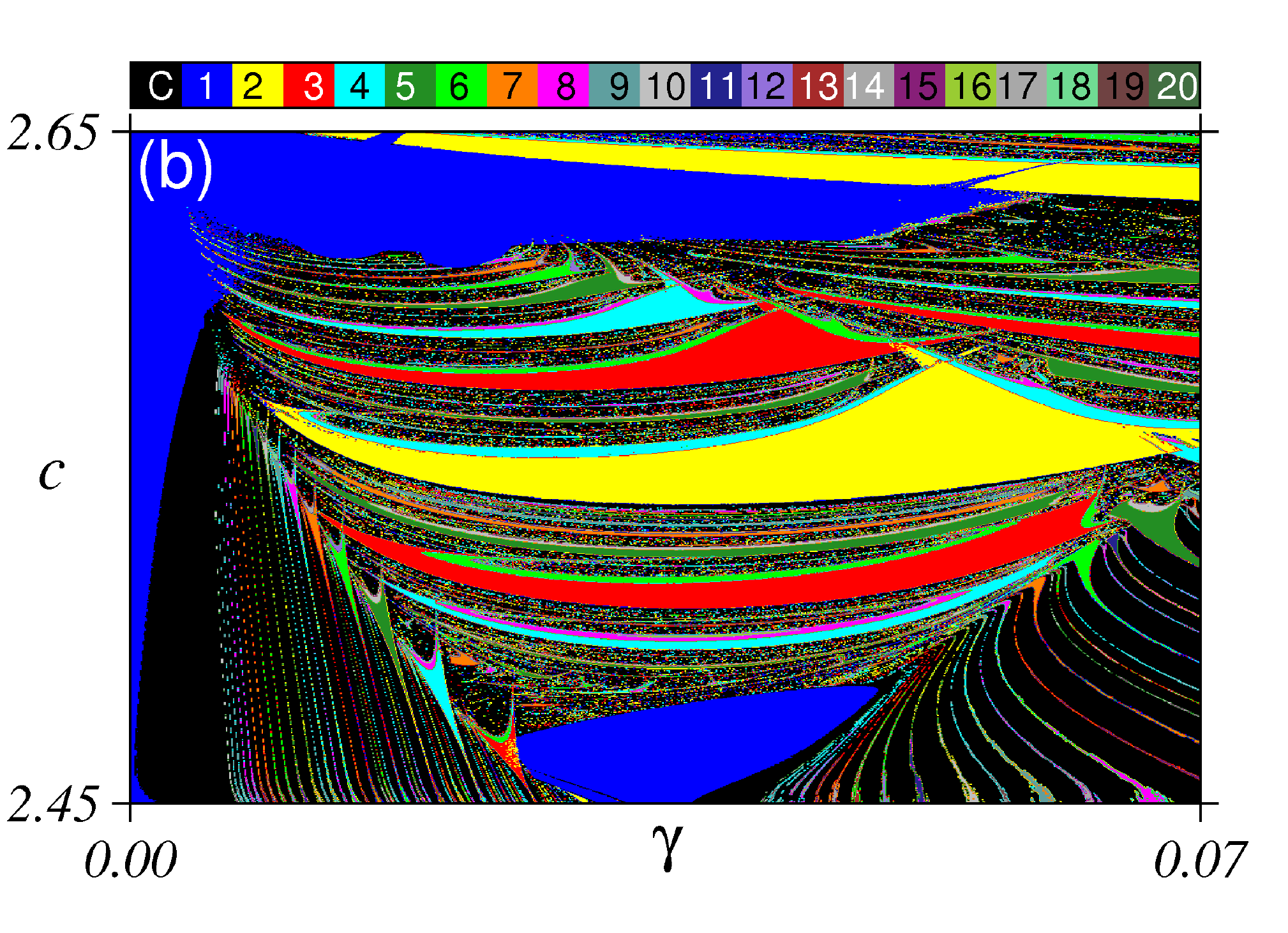}
  \caption{Unidirectional coupling case: (a) Lyapunov diagram for the
  largest exponents with bifurcation
  curves overlapped, and (b) isoperiodic diagram, both enlarged of
  Fig.~\ref{fig2}(c). In (a) the cyan and blue curves are saddle-node
  bifurcation curves, and the green curves are Neimark-Sacker
  bifurcation curves. In (b) the top color bar codifies the periods.
  Black corresponds to quasi-periodic/chaotic behaviors or periods
  greater than $20$.}
  \label{fig6}
\end{figure*}

\begin{figure*}[htb]
  \centering
  \includegraphics*[width=0.45\linewidth]{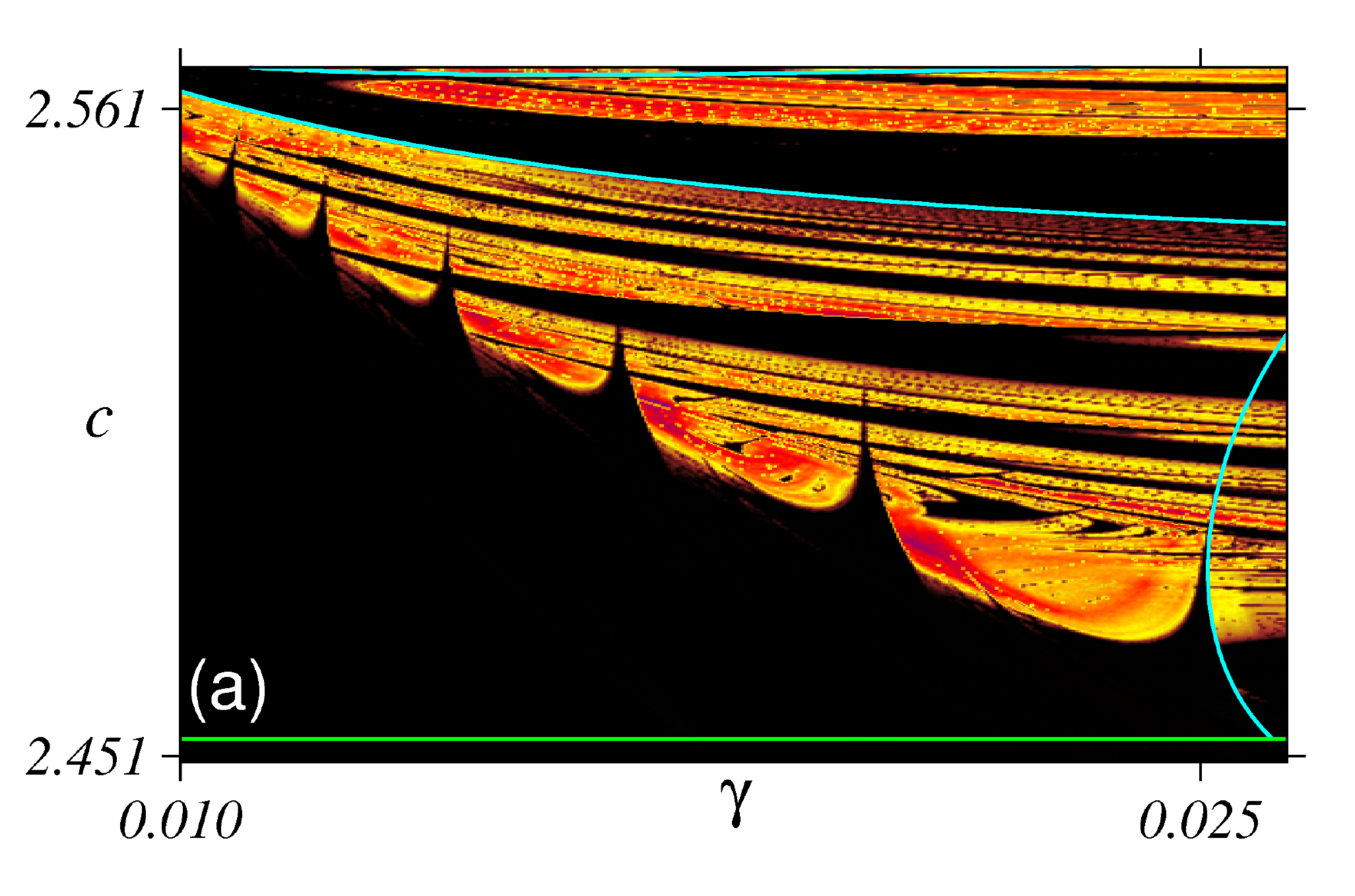}
  \includegraphics*[width=0.45\linewidth]{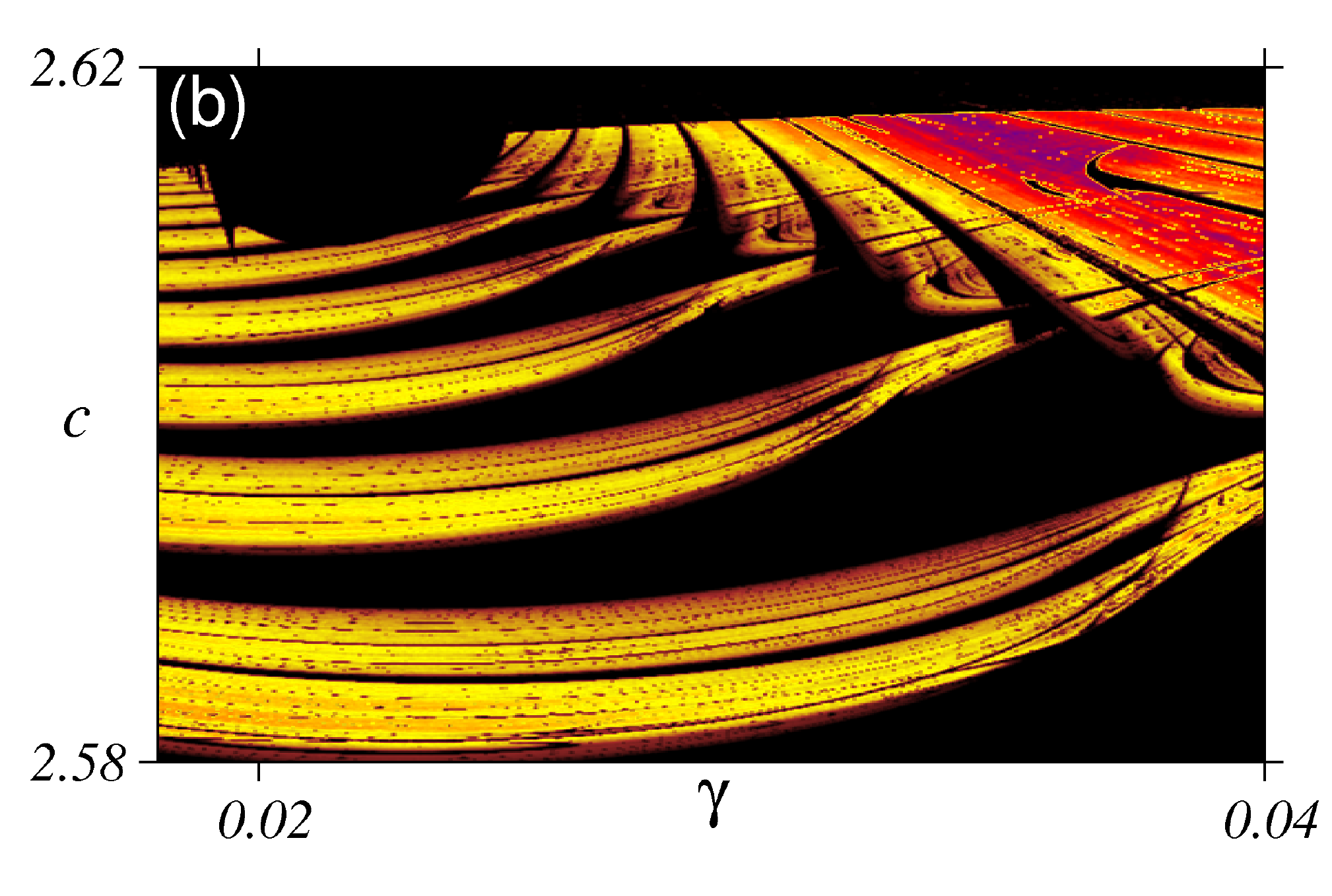}
  \includegraphics*[width=0.45\linewidth]{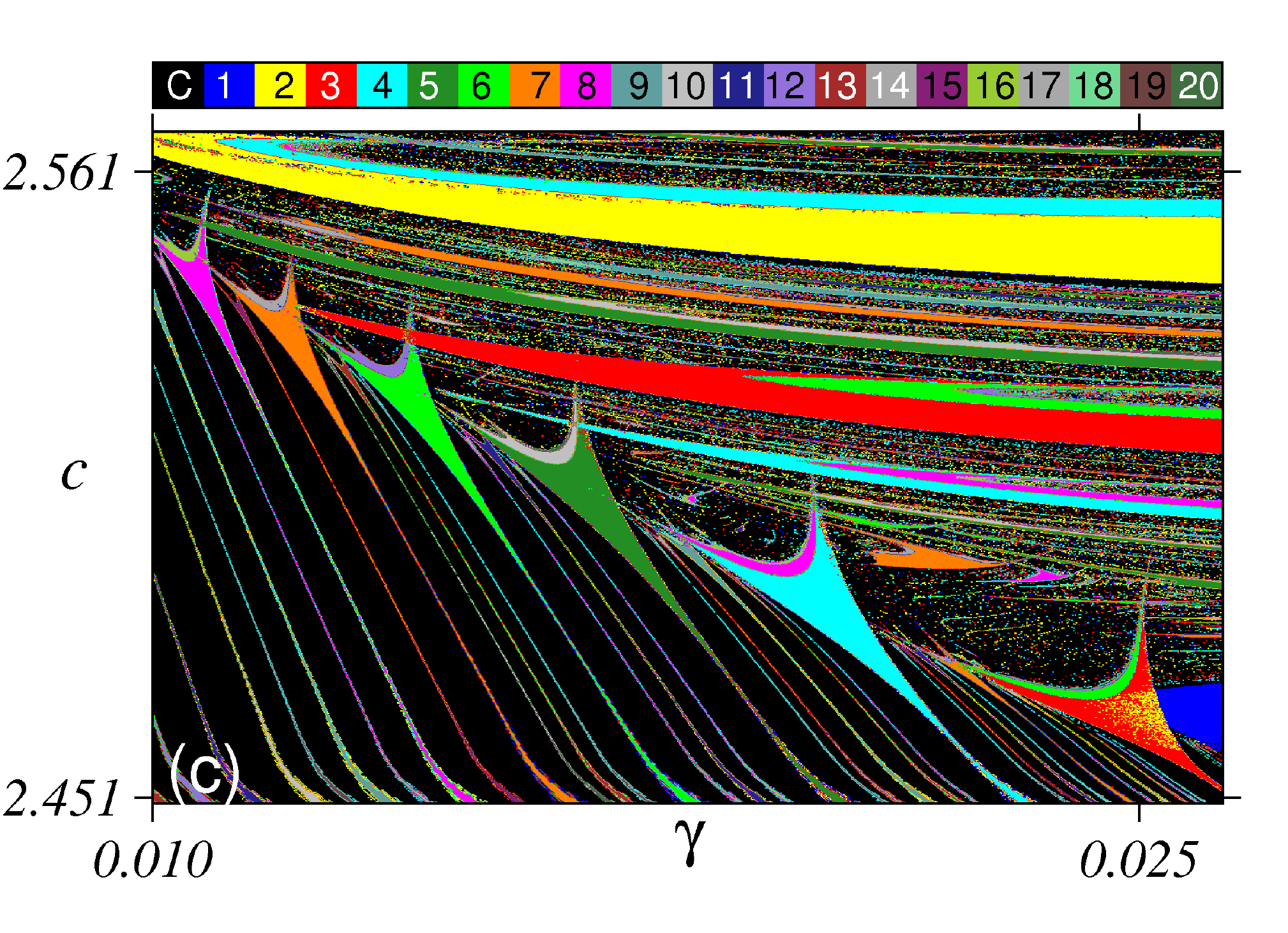}
  \includegraphics*[width=0.45\linewidth]{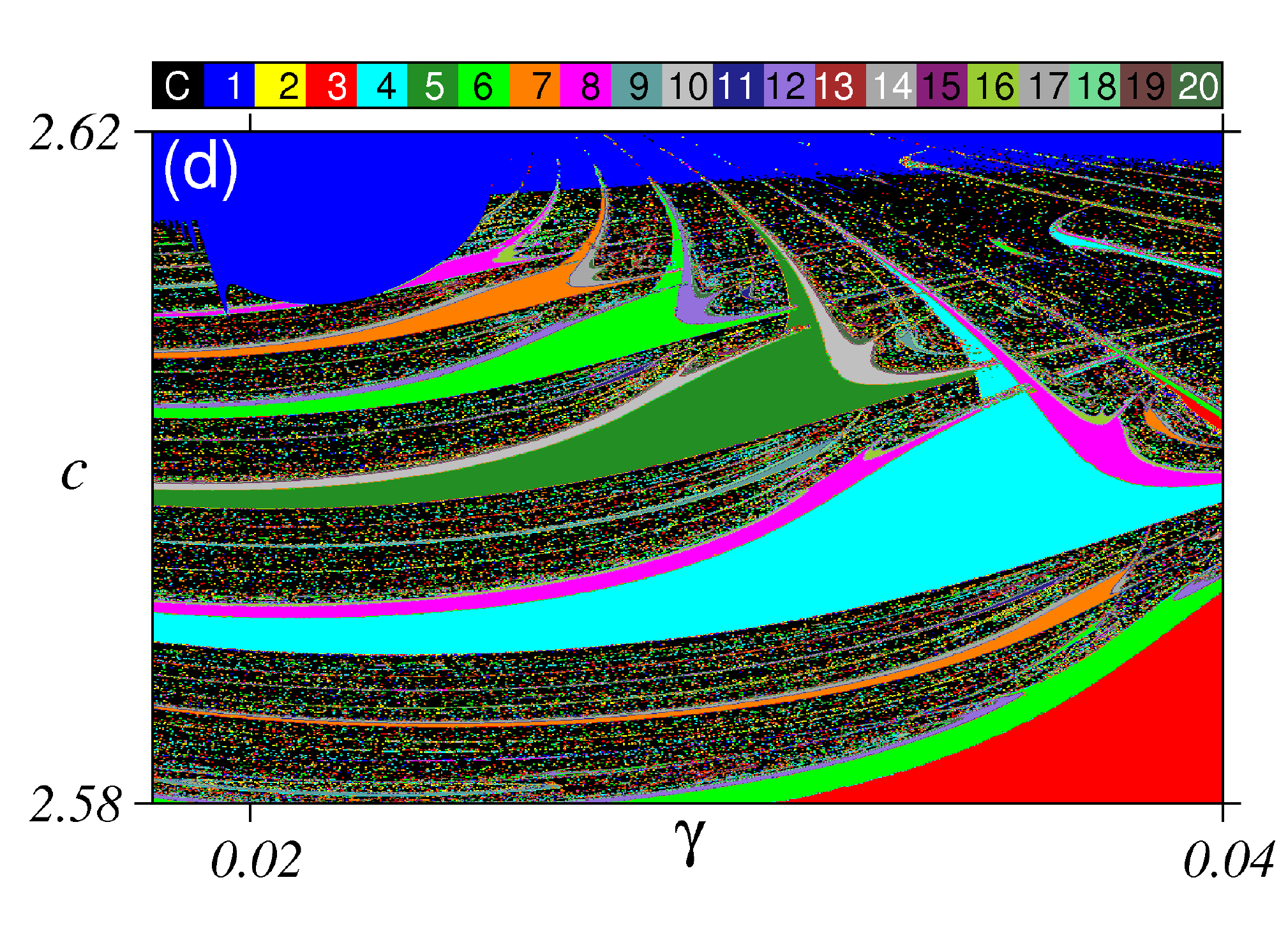}
  \caption{Unidirectional coupling case: (a) and (b) Lyapunov diagrams
  for the largest exponents, and
  (c) and (d) isoperiodic diagrams, both for the box B and C, respectively,
  of Fig.~\ref{fig6}. In (a) and (b) the colors follow the same of Fig.~\ref{fig2},
  and in (c) and (d) the top color bars codify the periods. Black corresponds
  to quasi-periodic/chaotic behaviors or periods greater than $20$.}
  \label{fig7}
\end{figure*}

\begin{figure*}[htb]
  \centering
  \includegraphics*[width=0.45\linewidth]{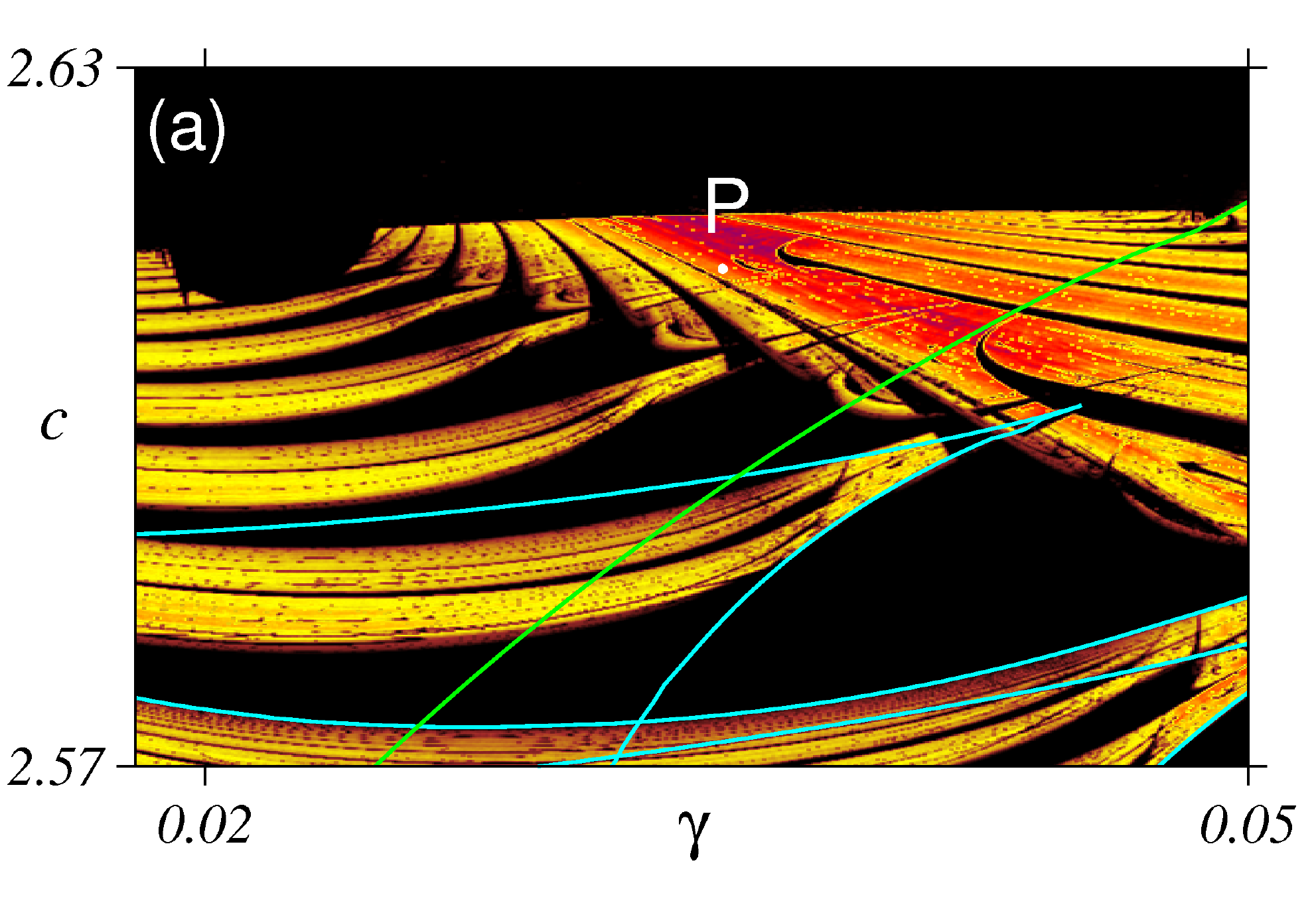}
  \includegraphics*[width=0.45\linewidth]{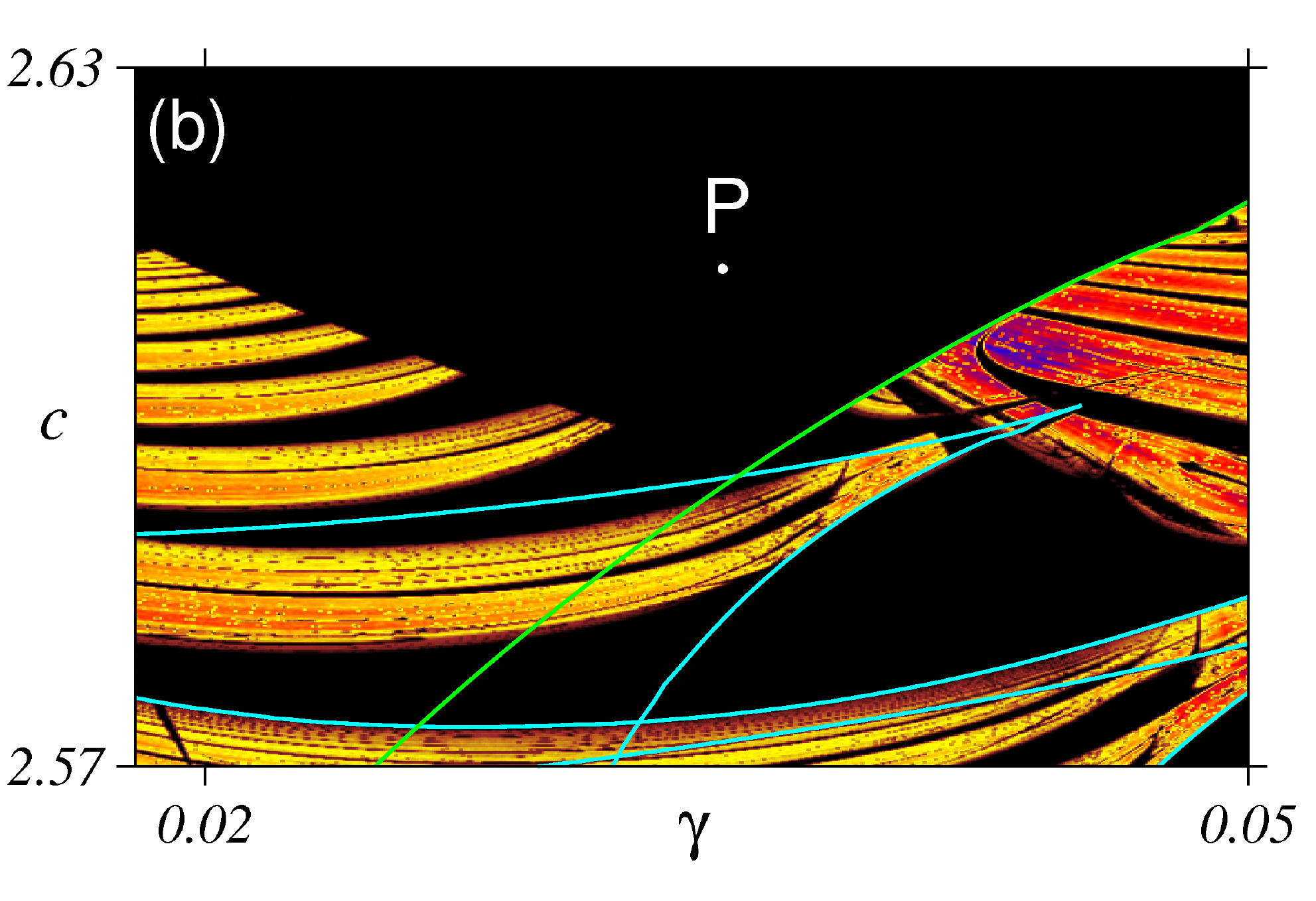}
  \includegraphics*[width=0.45\linewidth]{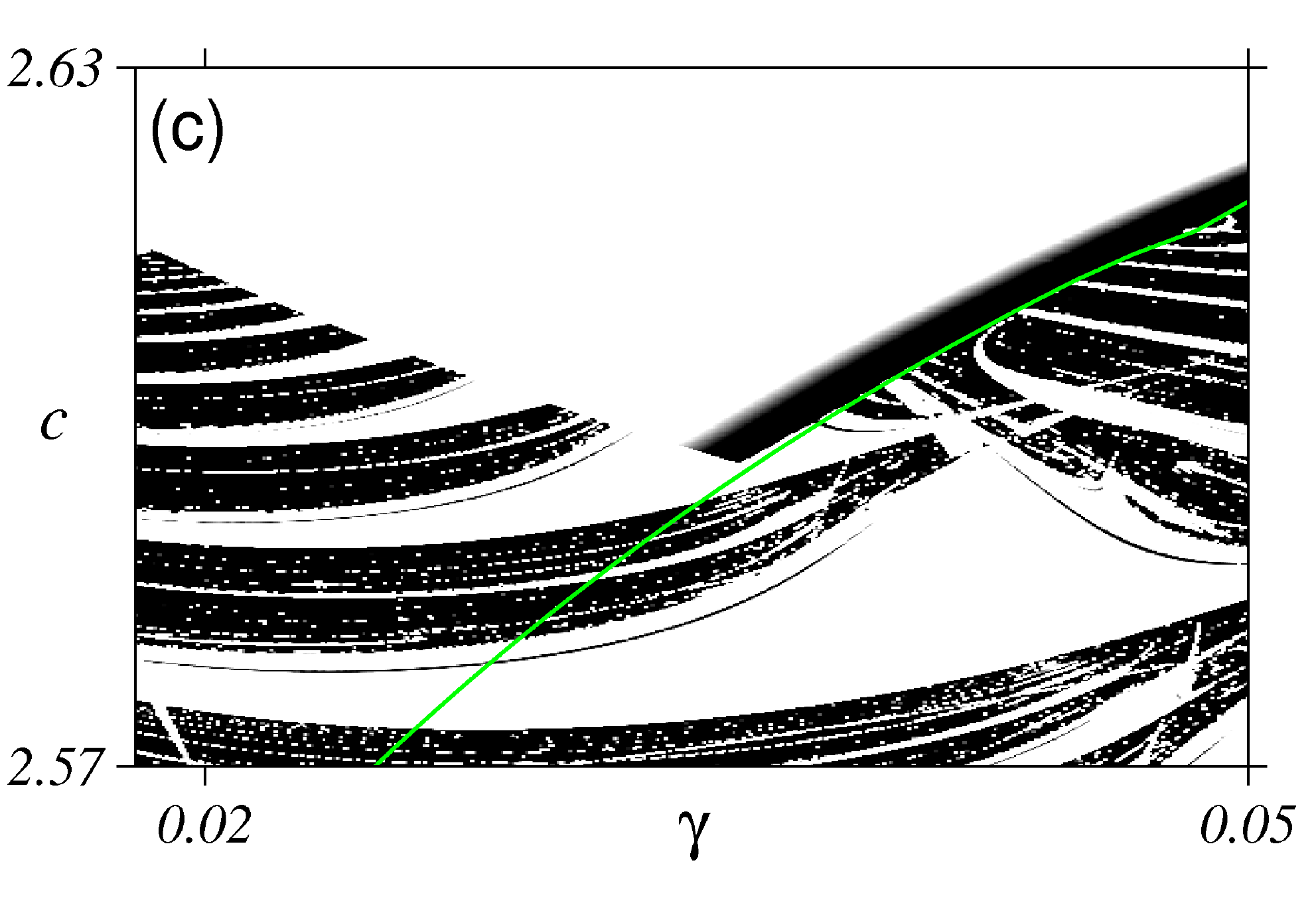}
  \includegraphics*[width=0.45\linewidth]{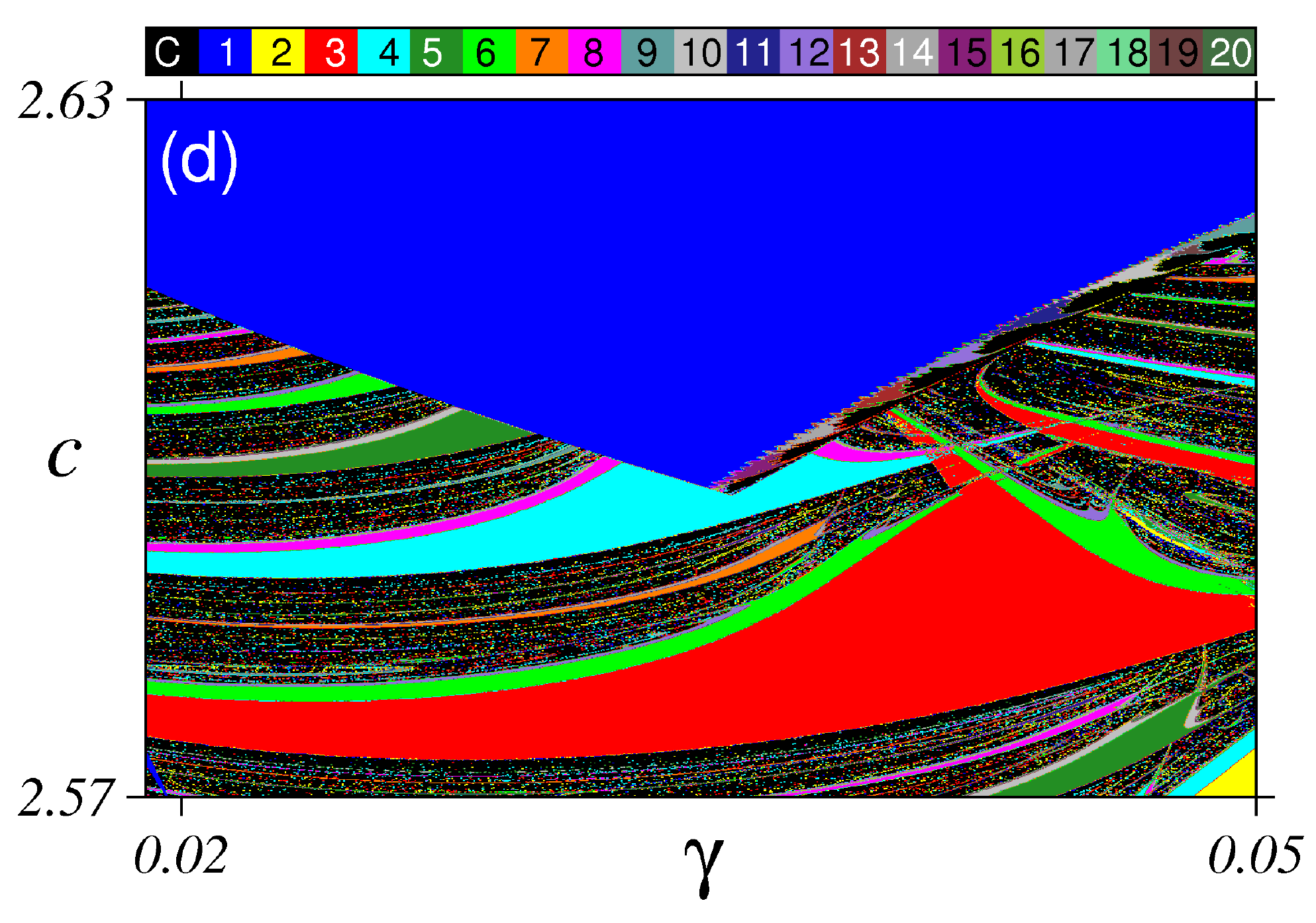}
  \caption{Unidirectional coupling case: (a) and (b) Lyapunov diagrams
    for the $\gamma \times c$
    plane for the largest exponents for two distinct initial
    conditions, and in (c) Lyapunov diagram for the second largest
    exponent for the same initial condition in (b). Cyan and green
    curves are saddle-node and Neimark-Sacker bifurcations,
    respectively.}
  \label{fig8}
\end{figure*}
\begin{figure*}[htb]
  \centering
  \includegraphics*[width=0.90\columnwidth]{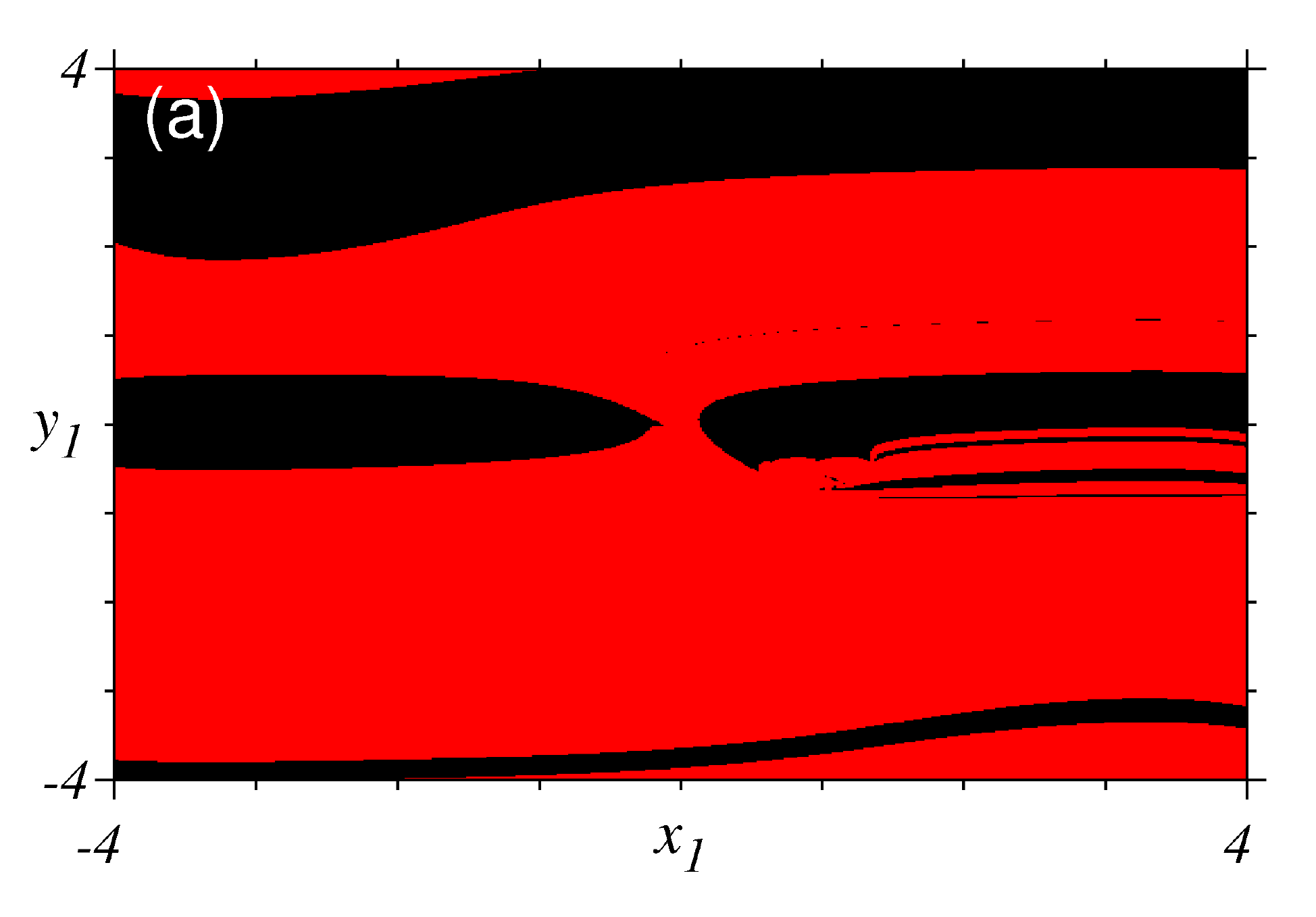}
  \includegraphics*[width=0.90\columnwidth]{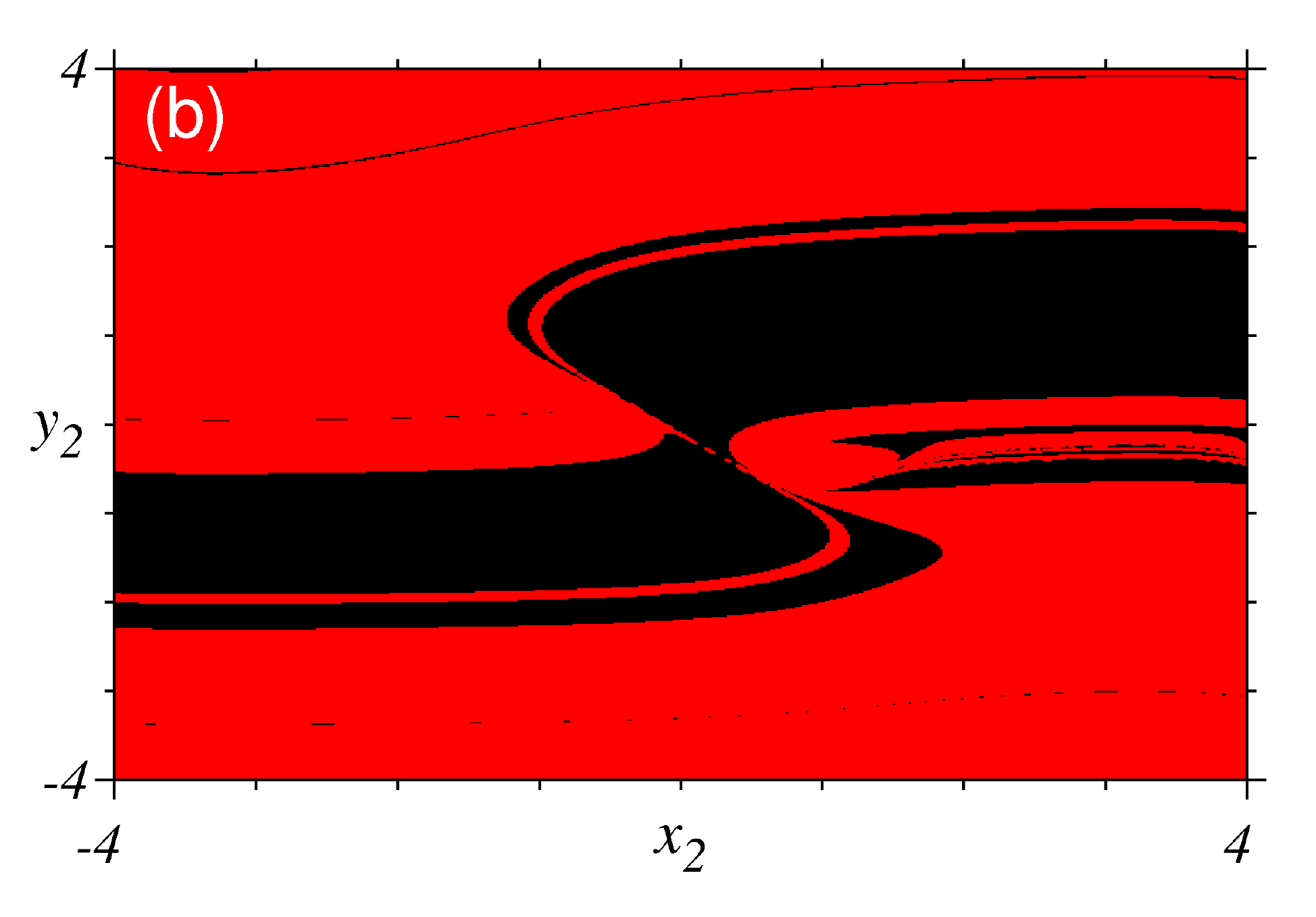}
  \caption{(a) and (b) planes of the four-dimensional basin of
  attractions for the point P in Figs.~\ref{fig8}(a) or (b). Black color
  is the basin of periodic attractors, and red one for chaotic attractors.}
  \label{fig9}
\end{figure*}

To corroborate the differences between the three distinct behaviors presented in
Figs.~\ref{fig4}(a) and (c), we show in Fig.~\ref{fig5} three attractors in three different
points of the Lyapunov diagram of Fig.~\ref{fig4}(a) (see the points I, II, and III).
In each plot we also show the respective Lyapunov spectrum (see the insets in Fig.~\ref{fig5}).

Figures~\ref{fig4}(b) and (d) show the amplification of the box B in Fig.~\ref{fig2}(a), and
another pattern formation of periodic structures can be visualized.
Here, the structures self-organizing in a circular layers pattern,
in which each layer is composed by two connected structure with same period.
A period-adding cascade occurs in direction of the inner layers.

Figures~\ref{fig2}(b) and (c) show the two other planes, (b) $\gamma \times b$,
and (c) $\gamma \times c$ for the largest Lyapunov exponents of the system~(\ref{eqFHN01}).
In Fig.~\ref{fig2}(b) the similarities with the $\gamma \times a$ plane (see Fig.~\ref{fig2}(a)),
are evident with the two types of organization patterns of structures shown in
Fig.~\ref{fig4}. In $\gamma \times c$ plane (see Fig.~\ref{fig2}(c)),
we observe the same periodic structures similar to the Arnold's tongues observed
previously in the $\gamma \times a$ plane (see Fig.~\ref{fig4}). For example,
Figs.~\ref{fig6}(a) and (b) are amplifications of the Lyapunov and
isoperiodic diagrams, respectively, for the box in Fig.~\ref{fig2}(c).
Bifurcation curves are also overlapped in the Lyapunov diagram of Fig.~\ref{fig6}(a),
where the cyan and blue curves are saddle-node bifurcations, and the green curves are
Neimark-Sacker bifurcation curves. It is clear to observe the same organization
pattern presented in Fig.~\ref{fig4}(a) and (c), quasi-periodic regions
(lower left and right portions of the diagram) with branches
of periodic structures embedded on it, and other branches of these structures
embedded on chaotic regions (higher portion of the diagram). The same
organization rule of periods of these structures was also observed. To corroborate
this statement, in Fig.~\ref{fig7}, we show amplifications of Fig.~\ref{fig6} and we
observe similarities with Figs.~\ref{fig4}(a) and (c). In Figs.~\ref{fig7}(b) and
(d) the same organization rule of the Arnold tongues in Figs.~\ref{fig7}(a) and (c),
or in Figs.~\ref{fig4}(a) and (c), is observed for the periodic structures of
Figs.~\ref{fig7}(b) and (d).

Another remarkable result is the agreement between the bifurcation curves in
Fig.~\ref{fig6}(a) and the periodic structures. In addition, there exist a
connection between the lower left Arnold tongues and the central periodic structures.
Initially, in Fig.~\ref{fig6}(a) it is clear to see two saddle-node bifurcation
curves (cyan curves) bordering two central periodic structures (the central yellow and
red periodic structures in Fig.~\ref{fig6}(b)) at their lower portions. Those
saddle-node bifurcations occur when, as the control parameters are varied, the
chaotic attractors become unstable and born stable periodic attractors.
Those features are corroborated by the isoperiodic
diagram in Fig.~\ref{fig6}(b). Following the second cyan curve (the curve that
border the red structure in Fig.~\ref{fig6}(b)) from right to left, we see that
after to border the lower portion of the periodic structure this curve reach the
respective left Arnold tongue of same period (the red Arnold tongue in Fig.~\ref{fig6}(b)).
Due to multistability the yellow Arnold tongue was hidden by the lower blue
periodic region (see Fig.~\ref{fig6}(b)). Moreover, due to numerical precision
to obtain those bifurcation curves, how smaller is the periodic structure, more
difficult is to obtain the curves. Therefore, for period-$4$ structure and on,
we have not accuracy to numerically obtain their saddle-node bifurcation curves.
However, our results show that the formation of the central set of periodic
structures is connected with the formation of the left set of Arnold tongues.

The blue saddle-node bifurcation curve in Fig.~\ref{fig6}(a) delimits the blue
periodic region in the bottom of the isoperiodic diagram of Fig.~\ref{fig6}(b).
The horizontal green Neimark-Sacker curve in Fig.~\ref{fig6}(a) (in the bottom)
shows the line where born the Arnold tongues, corroborated by the isoperiodic
diagram in Fig.~\ref{fig6}(b), where in the bottom of the diagram we see the
branches of the Arnold tongues in the quasiperiodic domains (lower black regions
in Fig.~\ref{fig6}(b)). The diagonal green Neimark-Sacker curve in
Fig.~\ref{fig6}(a) has a strict connection with the multistability presented by
the system~(\ref{eqFHN01}) and with a family of Torus that emerges very close
to the curve. To show this feature, in Figs.~\ref{fig8}(a) and (b) we present
in more details a zoom of Fig.~\ref{fig7}(b), with the green Neimark-Sacker
and cyan saddle-node bifurcation curves for two different initial conditions:
(a) $(x_1, y_1, x_2, y_2) = (-0.1,0.1,0.5,-0.3)$, and
(b) $(x_1, y_1, x_2, y_2) = (-0.1,0.5,0.1,-0.3)$. By contrasting the diagrams
in (a) and (b) we see the effect of the initial condition and the role of the
green Neimark-Sacker curve, for different initial conditions the black
periodic border increases until near the limit of the green curve destroying
chaotic regions and periodic structures, and very close to the green curve
a Tori domain emerge. The Tori domain is observed in Fig.~\ref{fig8}(c) by plotting the
Lyapunov diagram with the second largest exponent. In this case, the
periodic domains are in white, Tori (quasiperiodicity) and chaotic ones are
in black. The Tori domain is the black strip just above the Neimark-Sacker bifurcation
curve (green curve), once that this strip is black in the diagrams of
Figs.~\ref{fig8}(b) and (c). To corroborate, in Fig.~\ref{fig8}(d) we present the
isoperiodic diagram of Fig.~\ref{fig8}(b). In that diagram, near to the border
of the blue region emerge Arnold tongues and a Tori domain close to the
Neimark-Sacker curve, delimiting the boundary between the Tori and chaotic domains.

In Figs.~\ref{fig9}(a) and (b) we show two planes of the
four-dimensional basin of attraction of the point P in Figs.~\ref{fig8}(a) and (b).
The basins are constructed using the largest Lyapunov exponent and
associating black and red colors for periodic and chaotic attractors, respectively.
The initial conditions in Fig.~\ref{fig9}(a) are $(x_2, y_2) = (0.5,-0.3)$, and in
Fig.~\ref{fig9}(b) $(x_1, y_1) = (-0.1,0.1)$. It is easy to observe the existence of large
periodic domains (black regions) in the basins of attraction.

Regarding the two coupled FHN oscillators with bidirectional coupling,
Figs.~\ref{fig2}(d)-(f), the dynamics does not display the richness of details
shown in Figs.~\ref{fig2}(a)-(c), for the two coupled FHN oscillators with
unidirectional coupling. Comparing the dynamics, presented in the Lyapunov diagrams
of Fig.~\ref{fig2}, for both coupling, it is possible
to conclude that the unidirectional coupling presents a more subtlety in the dynamics
concerning the presence of self-organized periodic structures embedded in quasi-periodic
and chaotic regions in the three planes of parameters, namely $\gamma \times a$,
$\gamma \times b$, and $\gamma \times c$.

\section{Conclusions}
\label{conc}

In this paper we have investigated the dynamics of FHN oscillators
composed by two coupled models, using unidirectional and
bidirectional coupling. Each oscillator is modeled by a set of two
autonomous nonlinear first-order ordinary differential equations
that describes the dynamics of a nerve impulse through the
neuronal membrane, namely FitzHugh-Nagumo model. Our main goal is
the investigation of the influence of the coupling strength between
the oscillators with the type of coupling. For this purpose, we constructed the Lyapunov and
isoperiodic diagrams, which are parameter-planes with the Lyapunov
exponents and periods codified by colors, respectively, for the
two cases of the FHN-networks. By contrasting those diagrams in each case,
we observe the changes in the dynamical
behaviors of these models. We showed that $\gamma$
also has an important role in the dynamics of the FHN-networks.
In addition, we have obtained the bifurcation
curves by numerical continuation method, and overlapped those curves in the
Lyapunov diagrams, unveiling the existence of a connection between different
types of periodic structures through those bifurcation curves.

Regarding the two coupled FHN oscillators for the unidirectional coupling,
{\it i.e.}, the coupling only occurs in one of the four variables
$(x_i, y_i; i = 1, 2)$, namely $x_1$ in system~(\ref{eqFHN01}). The dynamics
for this case in the Lyapunov and isoperiodic diagrams presents the
coexistence of periodic, quasi-periodic, and chaotic oscillations,
and the presence of stability domains (periodic structures) embedded
in quasi-periodic and chaotic domains. The dynamical behavior presented
in this system resembles with the dynamics of systems that show the presence of Arnold tongues.
The organization rule of periods, shown in Figs.~\ref{fig4}(b)
and \ref{fig6}(c), seems to be a branch of the Stern-Brocot tree previously
reported in a three-dimensional version of the FHN model \cite{gal1}.
A circular organization of periodic structures in layers was also observed in a portion
of the parameter-plane (Figs.~\ref{fig4}(b) and (d)).

For the bidirectional
case, the coupling occurs in two of the four variables $(x_i, y_i; i = 1, 2)$,
namely $x_1$ and $x_2$ in system~(\ref{eqFHN3}). In this case,
the dynamics change drastically with the disappearance of the
quasi-periodic domain and of the Arnold tongue-like structures,
surviving the chaotic domains and some periodic structures (see Fig.~\ref{fig2}).
Among some examples of discrete- and continuous-time systems
that present Arnold tongues in the parameter-planes~\cite{schus,rech2,podh},
forced oscillators show a large spectrum of applications and Arnold tongues
are abundantly observed in those systems. In our work we characterize
the Arnold tongues in the parameter-planes of the unidirectional
coupling case that, indeed, can be interpreted as a forced oscillator, being
$(x_2, y_2)$ the master and $(x_1, y_1)$ the slave in system~(\ref{eqFHN01}).
The primary Arnold tongues and periodic structures shown in the
isoperiodic diagrams of Figs.~\ref{fig4}(c),~\ref{fig6},~\ref{fig7},
and~\ref{fig8}, are organized by period-adding cascades with
period-$1$ being the factor of the adding. An interesting
verification is that the master oscillates in cicle-$1$ driving
the slave with a period-$1$ oscillations, which is the factor of the
period-adding cascades of the slave.

For networks of two models, and for both coupling cases,
the Lyapunov diagrams are slightly different over wide regions (see
Fig.~\ref{fig2}) but a
general feature was observed in some regions of these diagrams,
namely the existence of periodic structures embedded in chaotic
regions. These sets of periodic structures are presented in a wide
range of nonlinear systems~\cite{alan,manc,gal4,stoop}. An exception is in
high-dimensional systems with more than three-dimensions, where
hyperchaotic behaviors can occur. In these systems, on the
hyperchaotic regions there are no periodic structures or the
structures are malformed or shapeless~\cite{cris}.

Relating the study presented here with the synchronization regimes
in the two coupled FHN systems, it was also observed that, for both
systems (unidirectional and bidirectional coupling) initializing in
the same initial conditions, lead to synchronized states. On the other
hand, initializing both systems (unidirectional and bidirectional
coupling) in different initial conditions, lead to non-syn\-chronized
states in the periodic domains of the parameter-planes of Fig.~\ref{fig1}.
Moreover, chaotic synchronization was not observed in the chaotic domains
covered in our study.

\section*{Acknowledgments}

The authors thank Conselho Nacional de Desenvolvimento Cient\'\i
fico e Tecnol\'ogico-CNPq, Coordena\c c\~ao de Aperfei\c co\-amento de
Pessoal de N\'\i vel Superior-CAPES, Funda\c c\~ao de Amparo \`a
Pesquisa e Inova\c c\~ao do Estado de Santa
Catarina-FAPESC, Brazilian agencies, for financial support.
%

\end{document}